\documentclass[12pt]{iopart}
\usepackage{amssymb}
\usepackage{epsfig,color}

\begin{document}

\title[F Kh Abdullaev \textit{et al}]{Bright solitons in Bose-Einstein
condensates with field-induced dipole moments}
\author{F Kh Abdullaev$^{1,2}$, A Gammal$^{3}$, B A Malomed$^{4}$ and Lauro
Tomio$^{1,5}$}
\address{
$^1$Instituto de F\'\i sica Te\'orica, UNESP-Universidade Estadual Paulista,
01140-070, S\~ao Paulo, Brazil.\\
$^2$Department of Physics, Kulliyyah of Science, International Islamic
University of Malaysia, 25200, Kuantan, Malaysia.\\
$^3$Instituto de F\'\i sica, Universidade de S\~ao Paulo, 05508-090, S\~ao
Paulo, Brazil.\\
$^4$Department of Physical Electronics, School of Electrical Engineering,
Faculty of Engineering, Tel Aviv University, Tel Aviv 69978, Israel.\\
$^5$Centro de Ci\^encias Naturais e Humanas (CCNH), Universidade Federal do ABC,
09210-170, Santo Andr\'e, Brazil.}

\date{\today}

\begin{abstract}
We introduce an effectively one-dimensional (1D) model of a bosonic gas of
particles carrying collinear dipole moments which are induced by an external
polarizing field with the strength periodically modulated along the
coordinate, which gives rise to an effective \textit{nonlocal nonlinear
lattice} in the condensate. The existence, shape and stability of bright
solitons, appearing in this model, are investigated by means of the
variational approximation and numerical methods. The mobility of solitons
and interactions between them are studied too.
\end{abstract}


\maketitle

\section{Introduction}

Ultracold bosonic gases with dipole-dipole interactions (DDI) have drawn a
great deal of attention in the last years, which is stimulated by new
experimental achievements in the achievement of the Bose-Einstein
condensation (BEC) in gases made of atoms carrying permanent magnetic
moments, such as chromium \cite{Cr}, dysprosium \cite{Dy}, and erbium 
\cite{Er} , and the development of efficient theoretical methods for the analysis
of such condensates \cite{Oxford}. The long-range and anisotropic character
of the DDI leads to new physical phenomena, which are not expected in BEC
with contact interactions, see reviews~\cite{Baranov,Lahaye,Carr}. Among
these phenomena well known are pattern-formation scenarios \cite{patt-form},
the $d$-wave collapse \cite{d-wave}, nonlocally coupled solitons in stacked
systems \cite{stack}, multidimensional anisotropic solitons \cite{Tikho},
solitons in dipolar condensates in optical lattices~\cite{Cuevas,Luis,Gligori,SKA}, 
and others. In addition to direct-current (dc)
external magnetic fields, various configurations of dipolar condensates can
be also controlled by combinations including alternating-current (ac)
components \cite{Michelli,Raymond}.

Another novel and potentially important ingredient available in BEC settings
is spatially periodic modulation of the local strength of the contact
interactions by means of the Feshbach resonance controlled by periodically
patterned laser or magnetic fields, as considered, e.g., in \cite{3Dcross},
leading to the concept of optically or magnetically induced \textit{nonlinear 
lattices}. The nonlinearity modulation in space gives rise to new
types of solitons and solitary vortices, as summarized in review~\cite{KMT}.
In particular, it was recently demonstrated that periodic modulation of the
local orientation of permanent atomic or molecular dipole moments in an
effectively one-dimensional (1D) setting, which may be induced by a
periodically inhomogeneous external polarizing field, makes it possible to
create DDI-induced \textit{nonlocal} nonlinear lattices in atomic
condensates \cite{AGMT}. The necessary periodic field structure can be built
using the available technique of magnetic lattices \cite{magn-latt}, or
similar ferroelectric lattices (see,e.g., Ref. \cite{ferroelectric}). The
analysis in Ref. \cite{AGMT} was focused on 1D solitons in the dipolar
condensate with periodic variations of the angle between dipoles.

Another possibility for the creation of nonlocal nonlinear lattices and the
study of self-trapped matter-wave modes in them is to use a bosonic gas of
polarizable particles, which do not carry permanent dipole moments, while a
spatially periodic distribution of the dipolar density is \emph{induced} by
an external spatially varying \emph{polarizing field} \cite{Raymond}. In
particular, a promising possibility is to consider an ultracold gas of
polarizable molecules, with dipole moments induced by a spatially modulated
dc electric field (such periodic settings were not considered in \cite{Raymond}). 
The DDI in gases of field-induced dipoles may be very strong and
give rise to a number of significant effects~\cite{Lahaye,induced,Raymond}.

In the present work we consider a quasi-1D dipolar BEC with electric dipole
moments of particles induced by the dc field with the local strength
periodically varying in space, while its direction is uniform, being
oriented along the system's axis, thus giving rise to the attractive DDI in
the condensate (on the contrary to the repulsive interactions between
induced dipoles polarized perpendicular to the system's axis or plane, which
was the subject of the analysis in \cite{Raymond}). The objective is to
investigate the existence and stability of bright solitons in this system,
controlled by the effective DDI-induced nonlocal nonlinear lattice. The
mobility and collisions of the solitons are considered too.

The paper is organized as follows. The models is presented in Sec. II, which
is followed by the study of the existence and stability of bright solitons
in Sec. III. Results for dynamics and collision of solitons are reported in
Sec. IV. Conclusions and perspectives are summarized in Sec. V.

\section{The model}

We consider the condensate elongated along axis $x$, with dipole moments of
polarizable molecules or atoms induced by an external field directed along 
$x$ too. The local strength of the polarizing field also varies along $x$.
Accordingly, one can use the effectively one-dimensional (1D)
Gross-Pitaevskii equation (GPE), with the DDI term derived from the
underlying 3D GPE, as shown below. The necessary spatially modulated dc
electric and/or magnetic field can be imposed, as said above, by
ferroelectric or ferromagnetic lattices. Below, we consider local dipole
moment induced by a polarizing electric field. One can also consider
magnetic dipole moments induced by a solenoid, as shown at the end of this
section.

In addition to the ferroelectric lattice, the periodic modulation of the
strength of the electric field oriented \emph{perpendicular} to the system's
axis ($x$) can be provided by a capacitor with the separation between its
electrodes modulated in $x$ periodically, as per Eq. (\ref{D}) written
below, cf. Ref. \cite{Raymond}. However, the most essential part of the
analysis is developed below for the periodically modulated strength of the
electric field directed \emph{along} $x$, to fix the attractive character of
the respective DDI. This configuration of the electric field can be provided
by a stacked capacitor built along the $x$ axis (with the array of parallel
electrodes made as grids, to prevent interference with the BEC flow along
the axis), assuming periodic modulation of the dc voltage applied to
adjacent pairs of electrodes.

\subsection*{\textbf{The derivation of the 1D Gross-Pitaevskii equation}}

The GPE for the 3D mean-field wave function $\Psi (\mathbf{r},t)$ is
\begin{eqnarray}
\hspace{-1cm}\mathrm{i}\hbar \frac{\partial \Psi }{\partial t} &=&-\frac{
\hbar ^{2}}{2m}\nabla^2 \Psi +\frac{m}{2}\left[ \omega _{\parallel
}^{2}x^{2}+d(x)\mathcal{E}(x)+\omega _{\perp }^{2}(y^{2}+z^{2})\right] \Psi +
\nonumber \\
&+&g_{\mathrm{3D}}|\Psi |^{2}\Psi +\left[ \int |\Psi (\mathbf{r}^{\prime
},t)|^{2}W_{\mathrm{DD}}(\mathbf{r}-\mathbf{r}^{\prime })d^{3}\mathbf{r}
^{\prime }\right] \Psi ,  \label{eq3}
\end{eqnarray}
where $d(x)=\gamma \mathcal{E}(x)$ is the local dipole moment, induced by
external field $\mathcal{E}(x)$, which is directed and modulated along $x$, 
$\gamma $ is the molecular or atomic polarizability, $g_{\mathrm{3D}}$ is
defined by the two-body scattering length $a_{s}$ and atomic mass $m$: 
$g_{\mathrm{3D}}\equiv 4\pi \hbar ^{2}a_{s}/m$. Further, the DDI kernel is
given by 
\begin{equation}
W_{\mathrm{DD}}(\mathbf{r-r}^{\prime })=\frac{d(x)d(x^{\prime })}{\left\vert
\mathbf{r-r}^{\prime }\right\vert ^{3}}\left[ 1-\frac{3\left( x-x^{\prime
}\right) ^{2}}{\left\vert \mathbf{r-r}^{\prime }\right\vert ^{2}}\right]
\label{Vd}
\end{equation}
and the wave function is normalized to the number of atoms,
\begin{equation}
N=\int \left\vert \Psi (\mathbf{r},t)\right\vert ^{2}d^{3}\mathbf{r}.
\label{norm}
\end{equation}

The field-induced dipole moment is essential in the range of
\begin{equation}
d\cdot \mathcal{E}\sim B,  \label{B}
\end{equation}
where $B$ is the rotational constant, determined by to the equilibrium
internuclear distance $r$ and reduced mass $m_{r}$ of the polarizable
molecule: $B=\hbar ^{2}/(2m_{r}r^{2})$ \cite{Lahaye}. Typical values of the
parameters are: $d\sim $ 1 Debye, $B\sim h\times $ $10$ GHz, which yields an
estimate for the necessary electric-field strength, $\mathcal{E}\sim 10^{4}$
V$\cdot $cm$^{-1}$. Such fields are accessible to experiments with BEC in
atomic gases (see Appendix B of Ref. \cite{Lahaye}).

Thus, the spatial variation of the strength of the polarizing dc electric
field,
\begin{equation}
\mathcal{E}(x)=\mathcal{E}_{0}f(x),  \label{E}
\end{equation}
leads to the respective spatial modulation of the DDI, with $d(x)=d_{0}f(x)$. 
In particular, the periodic variation of the field, such as that adopted
below in Eq. (\ref{Phi}), induces the above-mentioned effective nonlocal
nonlinear lattice in the GPE.

To derive the equation for the wave function in the quasi-1D case, we use
the method elaborated in Ref. \cite{Sinha}. If the ground state in the
transverse plane, $\left( y,z\right) $, is imposed by the trapping
potential, the 3D wave function may be factorized as usual \cite{Luca}
\begin{equation}
\Psi (\mathbf{r},t)=\psi (x,t)\left( \sqrt{\pi }a_{\perp }\right) ^{-1}\exp
\left( -\rho ^{2}/2a_{\perp }^{2}\right) ,  \label{psi}
\end{equation}
with $\rho ^{2}\equiv y^{2}+z^{2}$, and $a_{\perp }^{2}\equiv \hbar /m\omega
_{\perp }$. Substituting this expression into (\ref{eq3}) and integrating
over $(y,z)$, the effective one-dimensional DDI is derived with kernel
\begin{equation}
W_{1\mathrm{DD}}=\frac{2d^{2}}{a_{\perp }^{3}}\left[ \frac{2|x|}{a_{\perp }}
-\sqrt{\pi }\left( 1+\frac{2x^{2}}{a_{\perp }^{2}}\right) \exp \left( \frac{
x^{2}}{a_{\perp }^{2}}\right) \mathrm{erfc}\left( \frac{|x|}{a_{\perp }}
\right) \right] ,  \label{V1D}
\end{equation}
where $\mathrm{erfc}$ is the complementary error function. Next, we
introduce dimensionless variables,
\[
\hspace{-1cm}x\rightarrow {a_{\perp }}x,\;\;t\rightarrow \frac{t}{\omega
_{\perp }},\;\;\psi (x,t)\rightarrow \sqrt{\frac{5}{\pi ^{3/2}a_{d}}}\phi
(x,t),\;\;\alpha =\frac{\omega _{\parallel }^{2}}{2\omega _{\perp }^{2}}
,\;\;g=\frac{10a_{s}}{\pi ^{3/2}a_{d}},
\]
where $a_{d}=md^{2}/\hbar ^{2}$ is the characteristic DDI\ length. Defining
the rescaled polarizability, $\beta \equiv \gamma \mathcal{E}_{0}^{2}/{
2(a_{\perp }\omega _{\perp })^{2}}$, where $\mathcal{E}_{0}$ is the
amplitude of modulated field (\ref{E}), the original 3D GPE (\ref{eq3}) is
thus reduced to the 1D equation:
\begin{equation}
\hspace{-1cm}\mathrm{i}\frac{\partial \phi }{\partial t}=-\frac{1}{2}\frac{
\partial ^{2}\phi }{\partial x^{2}}+\alpha x^{2}\phi +\beta f^{2}(x)\phi
+g|\phi |^{2}\phi -f(x)\phi \int_{-\infty }^{+\infty }f(x^{\prime })|\phi
^{\prime }|^{2}R(x-x^{\prime })dx^{\prime },  \label{eq1}
\end{equation}
where $\phi \equiv \phi (x,t)$ and $\phi ^{\prime }\equiv \phi (x^{\prime
},t)$, with the effective 1D~kernel, following from Eq. (\ref{V1D}) is (cf.
Ref. \cite{Sinha}), given by
\begin{equation}
R(x)=\sigma \frac{10}{\pi }\left[ \left( 1+2x^{2}\right) \exp \left(
x^{2}\right) \mathrm{erfc}(|x|)-\frac{2}{\sqrt{\pi }}|x|\right] .
\label{kernel}
\end{equation}
Here we have added an extra parameter, which takes values $\sigma =+1$ for
the attractive DDI, and $\sigma =-1/2$ for the repulsive DDI between dipoles
oriented perpendicular to $x$, which makes it possible to consider the
latter case too. Actually, the rather complex kernel (\ref{kernel}) can be
replaced by a simplified expression,
\begin{equation}
R(x)=\frac{10\sigma }{\pi \sqrt{\left( \pi \;x^{2}+1\right) ^{3}}},
\label{R}
\end{equation}
which is very close to the exact one (\ref{kernel})~\cite{Cuevas}, barring
the fact that expression (\ref{R}) is smooth near $x=0$, while its
counterpart (\ref{kernel}) has a cusp at this point.

Note that the DDI can be represented by a pseudopotential which includes a
contact-interaction (delta-functional) term \cite{Baranov,Bortolotti}. Then,
spatially modulated $d(x)$ may induce a position-depending part of the
contact interactions too. Such a combination of nonlinear local and dipolar
lattices may be a subject of interest for a separate investigation. However,
in the present setting, the regularization scale $a_{\perp }$ in kernel (\ref
{V1D}) eliminates the singular part of the DDI at scales $|x|\lesssim
a_{\perp }$. Therefore, in the present work we restrict ourselves to the
consideration of the pure nonlinear nonlocal lattice.

We assume that the dynamics of the system in the perpendicular directions is
completely frozen, i.e., the transverse trapping frequency, $\omega _{\perp
} $, is much larger then the longitudinal one, $\omega _{\perp }\gg \omega
_{||}$. On the other hand, if $\omega _{\perp }$ is not too large,
interesting transverse effects may occur, such as the Einstein-de Haas
effect~\cite{Kawaguchi,Swislovski,Paz,Wall}, the consideration of which is
beyond the scope of the present work.

In the potential given in Eq. (\ref{eq1}) we can identify the usual
harmonic-trap potential, $\alpha x^{2}$, nonlinear term $g|\phi |^{2}\phi $
accounting for the collisional interaction, and an effective DDI potential,
composed of linear and a nonlinear terms:
\begin{equation}
V_{\mathrm{eff}}^{(\mathrm{DDI})}(x;|\phi |^{2})=f(x)\left[ \beta
f(x)-\int_{-\infty }^{+\infty }f(x^{\prime })|\phi ^{\prime
}|^{2}R(x-x^{\prime })dx^{\prime }\right] ,  \label{ddipot}
\end{equation}
where the modulation function [see Eq. (\ref{E})] is chosen, as said above,
in the form of a periodic one:
\begin{equation}
f(x)=f_{0}+f_{1}\cos (kx),  \label{eq2}
\end{equation}
with the constant parameters $f_{0}$, $f_{1}$, and $k\equiv 2\pi a_{\perp
}/\lambda =2\pi /\Lambda $. Parameter $\beta $ in Eq.(\ref{eq1}) can vary
from $1$ to $10$, under typical physical conditions, if the constant part of
the modulation function is fixed as $f_{0}\equiv 1$.

The Hamiltonian corresponding to Eq.~(\ref{eq1}) is
\begin{eqnarray}
H &=&\int_{-\infty }^{+\infty }dx\left[ \frac{1}{2}\left\vert \frac{\partial
\phi }{\partial x}\right\vert ^{2}+\frac{g}{2}|\phi |^{4}+\alpha x^{2}|\phi
|^{2}+\beta f^{2}(x)|\phi |^{2}\right]  \nonumber \\
&-&\frac{1}{2}\int_{-\infty }^{+\infty }dxf(x)|\phi |^{2}\int_{-\infty
}^{+\infty }dx^{\prime }f(x^{\prime })|\phi ^{\prime }|^{2}R(x-x^{\prime }).
\end{eqnarray}
Note that it contains not only the spatially modulated nonlinear DDI, but
also the additional linear potential, $\beta f^{2}(x)$, which is induced by
the interaction of the locally-induced dipole moment with the polarizing
field, cf. Ref. \cite{Raymond}. The Hamiltonian term corresponding to this
potential is denoted $H_{\mathrm{DE}}$ below.

\subsection*{\textbf{Evaluation of parameters}}

The energy of the interaction of the dipole with the external electric field
can be estimated, as per Eq. (\ref{B}), as $H_{\mathrm{DE}}=d\cdot \mathcal{E}
\sim B\sim h\times $ 10 GHz. As the total wave function is normalized to
the number of atoms [see Eq. (\ref{norm})], i.e., $|\Psi |^{2}\sim
N/a_{\perp }^{3}$, for the DDI\ energy we have
\begin{equation}
H_{\mathrm{DD}}=\int |\Psi |^{2}W_{\mathrm{DD}}(r)d^{3}r\sim {d^{2}}\frac{N}
{a_{\perp }^{3}}\sim \frac{a_{d}\hbar ^{2}}{m}\frac{N}{a_{\perp }^{3}},
\label{Hdd}
\end{equation}
where the characteristic DDI length is $a_{d}=d^{2}m/\hbar ^{2}$. Therefore,
with regard to $a_{\perp }^{2}=\hbar /(m\omega _{\perp })$, we obtain
\begin{equation}
H_{\mathrm{DD}}\sim N\left( a_{d}/a_{\perp }\right) \hbar \omega _{\perp }.
\label{Hdd2}
\end{equation}
For $N\sim 10^{5},\omega _{\perp }\sim 10^{4}$ Hz and $a_{d}\sim 10^{4}a_{s}$
, we thus conclude that $H_{\mathrm{DE}}\sim H_{\mathrm{DD}}$.

We note a peculiarity of the modulation period of the polarizing field in
the dimensionless equation. Physical values of the period should be,
evidently, on the order of or larger than the $\mathrm{\mu }$m scale. The
characteristic length related to $H_{\mathrm{DE}}$ is $l_{d}\sim \sqrt{\hbar
^{2}/mB}\sim 10^{-2}a_{\perp }$. Further, we use an estimate $B\sim {\hbar
^{2}/(ml_{d}^{2})}=10^{4}{\hbar ^{2}/(ma_{\perp }^{2})}=10^{4}{\hbar \omega
_{\perp }}$. Therefore, with $H_{\mathrm{DE}}\sim B$, we again obtain 
$H_{\mathrm{DE}}\sim H_{\mathrm{DD}}$. One possible way to suppress the
interaction represented by $H_{\mathrm{DE}}$, and thus to focus on nonlinear
DDI\ effects, is to decrease the induced dipolar moment $d$ up to the level
of $10^{-2}$ Debye (which remains experimentally observable \cite{Ni}), and
so to reduce $H_{\mathrm{DE}}$.

Another possible way is to suppress the linear-interaction term $H_{\mathrm{
\ DE}}$, which was proposed in Ref. \cite{Raymond}, is to consider the
polarization imposed by a combination of dc and ac electric fields 
\cite{Yi,Deb,McCarthy,Li,Golomedov}, oriented along the $z$- direction (i.e.,
perpendicular to the system's axis, $x$):
\begin{equation}
G(r)=F(r)[f_{\mathrm{dc}}+f_{\mathrm{ac}}\cos (\omega t)]\mathbf{e}_{z},
\label{Gr}
\end{equation}
Then the local dipolar moment $\mathbf{g}=g(t)\mathbf{e}_{z}$ of the atom or
molecule is determined by the intrinsic equation of motion, considered here
in the classical approximation \cite{Velde}:
\begin{equation}
\frac{d^{2}g}{dt^{2}}+\omega _{0}^{2}g+\Gamma \frac{dg}{dt}=F(r)[\lambda
(0)f_{\mathrm{dc}}+\lambda (\omega )f_{\mathrm{ac}}\cos (\omega t)],
\label{eq22}
\end{equation}
where $\omega _{0}$ is the intrinsic eigenfrequency, $\Gamma $ is the
damping coefficient, while $\lambda (0)\equiv \lambda _{0}$ and $\lambda
(\omega )$ are the effective static and dynamical susceptibilities,
respectively. In the off-resonance situation, when the ac frequency, $\omega
$, is not too close to $\omega _{0}$, the small dissipative term in 
Eq.~(\ref{eq22}) is negligible, which gives rise to the following solution:
\begin{equation}
g_{\mathrm{off}}(r)=F(r)\left[ \frac{\lambda _{0}}{\omega _{0}^{2}}f_{
\mathrm{dc}}+\frac{\lambda (\omega )f_{\mathrm{ac}}}{\omega ^{2}-\omega
_{0}^{2}}\cos (\omega t)\right] .  \label{goff}
\end{equation}

On the other hand, close to the resonance the ac drive yields:
\begin{equation}
g_{\mathrm{res}}(r)=F(r)\frac{\lambda (\omega _{0})}{\Gamma \omega _{0}}\sin
(\omega _{0}t).  \label{gres}
\end{equation}
These results lead to the following time-averaged DDI\ strength \cite{Raymond}:
\begin{eqnarray}
\langle g_{\mathrm{off}}(r_{1})g_{\mathrm{off}}(r_{2})\rangle
&=&F(r_{1})F(r_{2})\left[ \frac{\lambda _{0}^{2}}{\omega _{0}^{4}}f_{\mathrm{
\ dc}}^{2}+\frac{\lambda ^{2}(\omega )f_{\mathrm{ac}}^{2}}{2(\omega
_{0}^{2}-\omega ^{2})^{2}}\right] ,  \label{eq5} \\
\langle g_{\mathrm{res}}(r_{1})g_{\mathrm{res}}(r_{2})\rangle
&=&F(r_{1})F(r_{2})\frac{\lambda ^{2}}{2\Gamma ^{2}\omega _{0}^{2}}.
\nonumber
\end{eqnarray}
In addition to the DDI, in the off-resonance situation the field-induced
dipole moments give rise to the above-mentioned effective dipole-field
interaction:
\begin{equation}
V_{\mathrm{DE}}(r)=-\langle g_{\mathrm{off}}G\rangle =-\left( F(r)\right)
^{2}\left[ \frac{\lambda _{0}}{\omega _{0}^{2}}f_{\mathrm{dc}}^{2}+\frac{
\lambda (\omega )f_{\mathrm{ac}}^{2}}{2(\omega _{0}^{2}-\omega ^{2})}\right]
=-\chi F^{2}(r),  \label{eq7}
\end{equation}
where $\chi $ is the effective average polarizability, while in the resonant
situation, with $f_{\mathrm{dc}}=0$, the substitution of expression (\ref
{gres}) immediately yields $V_{\mathrm{DE}}(r)=0$. Then, it is obvious that
potential (\ref{eq7}) \textit{vanishes} at $\omega =\Omega $, with $\Omega $
defined by equation
\begin{equation}
\frac{\Omega }{\omega _{0}}=\sqrt{1+\frac{\lambda (\Omega )}{2\lambda (0)}
\frac{f_{\mathrm{dc}}^{2}}{f_{\mathrm{ac}}^{2}}},  \label{eq8}
\end{equation}
while the effective strength (\ref{eq5}) does not vanish under condition 
(\ref{eq8}):
\[
\langle g_{\mathrm{off}}(r_{1})g_{\mathrm{off}}(r_{2})\rangle
=F(r_{1})F(r_{2})\left[ \frac{\lambda _{0}^{2}f_{\mathrm{dc}}}{\omega
_{0}^{4}}\left( 1+2\frac{f_{\mathrm{dc}}^{2}}{f_{\mathrm{ac}}^{2}}\right)
\right] .
\]

{In the case when the local dipole moment is induced by magnetic field, we
consider the field with a fixed (}$x$) {orientation, produced by a solenoid
of diameter $D$, periodically varying along the solenoid's axis, $D=D(x)$,
with period $L$, such as
\begin{equation}
D(x)=D_{0}+D_{1}\cos \left( 2\pi x/L\right) .  \label{D}
\end{equation}
The local magnetic field in this configuration is
\begin{equation}
\hspace{-1cm}\mathcal{H}=\frac{\Phi }{\pi D^{2}(x)}=\Phi \left[
D_{0}+D_{1}\cos \left( \frac{2\pi x}{L}\right) \right] ^{-2}\approx \frac{
\Phi }{\pi D_{0}^{2}}\left[ 1-2\frac{D_{1}}{D_{0}}\cos \left( \frac{2\pi x}{L}
\right) \right] ,  \label{Phi}
\end{equation}
where $\Phi $ is the magnetic flux trapped in the solenoid, and the
approximation is valid for $D_{1}/D_{0}\ll 1$. It should be noted that this
scheme requires a strong magnetic field, which can be difficult to achieve
in the experiments. Therefore, in this work we actually consider induced
electric dipole moments. }

\section{Bright solitons: existence and stability}

The existence of bright-soliton solutions can be investigated by solving the
corresponding eigenvalue problem, obtained from Eqs.~(\ref{eq1}) and (\ref
{ddipot}), with $\phi = |\phi |e^{-\mathrm{i}\mu t}$:

\begin{equation}
-\frac{1}{2}\frac{\partial ^{2}\phi }{\partial x^{2}}+\alpha x^{2}\phi +
g|\phi |^{2}\phi + V_{eff}^{DDI}(x;|\phi|^2) \phi =\mu\phi .  \label{bseq}
\end{equation}
We consider full numerical solutions of Eq. (\ref{bseq}), as well as
corresponding variational approaches (VA), for two characteristic cases, as
defined in the next sub-sections. The VA results will be compared with the
numerical solutions.

\subsection{The variational approximation for $\protect\alpha =0$, $\protect
\beta \neq 0$.}

To derive the VA, we start from the averaged Lagrangian, $L=\int_{-\infty
}^{+\infty }\mathcal{L}dx,$ with density
\begin{equation}
\hspace{-1.5cm}\mathcal{L}=\mu |\phi |^{2}-\frac{1}{2}\left\vert \frac{d\phi
}{dx}\right\vert ^{2}-\beta \left[ f(x)\right] ^{2}|\phi |^{2}-\frac{g}{2}
|\phi |^{4}+\frac{f(x)}{2}|\phi |^{2}\int_{-\infty }^{+\infty }f(x^{\prime
})R(x-x^{\prime })|\phi ^{\prime }|^{2}dx^{\prime }.  \label{density}
\end{equation}
For the wave function of the condensate, we assume the following Gaussian
ansatz with center set at $x=\zeta $:
\begin{equation}
\phi =A\exp \left( -\frac{(x-\zeta )^{2}}{2a^{2}}\right) .  \label{gaussian}
\end{equation}
The corresponding averaged Lagrangian $L$ is given by
\begin{eqnarray*}
\frac{L}{N} &=&\mu _{r}-\frac{1}{4a^{2}}-\beta \left(
2f_{1}f_{0}e^{-a^{2}k^{2}/4}\cos (k\zeta )+\frac{f_{1}^{2}}{2}
e^{-a^{2}k^{2}}\cos (2k\zeta )\right) \\
&-&\frac{gN}{2\sqrt{2\pi }a}+\frac{N}{2\pi a^{2}}F(a,\zeta ,f_{0},f_{1}),
\end{eqnarray*}
where we have $N=\sqrt{\pi }A^{2}a$, $\mu _{r}\equiv \mu -\beta \left[
f_{0}^{2}+(1/2)f_{1}^{2}\right] $, and
\begin{equation}
\hspace{-1cm}F(a,\zeta ,f_{0},f_{1})\equiv \int_{-\infty }^{+\infty
}dxf(x)e^{-[(x-\zeta )/a]^{2}}\int_{-\infty }^{+\infty
}dyf(y)R(x-y)e^{-[(y-\zeta )/{a}]^{2}}.  \label{F}
\end{equation}
By means of a variable transformation, with $R$ defined by Eq.~(\ref{R}), 
$F$ can be represented in terms of single-variable integrals:
\begin{eqnarray}
F(a,\zeta ,f_{0},f_{1}) &=&\left[ f_{0}^{2}+\frac{1}{2}f_{1}^{2}\cos
{\ (2k\zeta )}e^{-(ka)^{2}/2}\right] h_{0}(a)  
\nonumber \\&+&
\frac{1}{2}f_{1}^{2}h_{2}(a)+2f_{0}f_{1}\cos {(k\zeta )}
e^{-(ka)^{2}/8}h_{1}(a),  \label{Fa}
\end{eqnarray}
where
\begin{eqnarray}
\left. h_{n}(a)\right\vert _{n=0,1,2} &\equiv &
\frac{10\sigma }{\sqrt{\pi }}a\int_{-\infty }^{+\infty }dz\frac{
e^{-(z/a)^{2}} }{\sqrt{2\pi \;z^{2}+1}^{3}} \cos \left( \frac{nkz}{\sqrt{2}}\right).  
\label{gn}
\end{eqnarray}
The corresponding Euler-Lagrange equations, $\partial {L}/\partial {N}=$ 
$\partial {\ L}/\partial {a}=$ $\partial {L}/\partial {\zeta }=0$, take the
following form:
\begin{eqnarray}
\mu &=&\frac{1}{4a^{2}}+ \left( {\sqrt{2\pi }ga}-{2F}\right) \frac{N}{2\pi a^{2}}
\nonumber \\ &+&
\beta \left[ f_{0}^{2}+\frac{f_{1}^{2}}{2}\left(
1+e^{-(ka)^{2}}\cos (2k\zeta )\right) +2f_{0}f_{1}e^{-(ka/2)^{2}}\cos
(k\zeta )\right],  \label{va-mu}
\\
N &=&2\pi \frac{1+2\beta (ka^{2})^{2}f_{1}\left[ f_{0}e^{-(ka/2)^{2}}\cos
(k\zeta )+f_{1}e^{-(ka)^{2}}\cos (2k\zeta )\right] }{4F-2a\partial {F}
/\partial {a}-\sqrt{2\pi }ga},  \label{va-N} \\ \nonumber \\
\zeta &=& 0,\;\;\pi /k, \;\;2\pi /k, ... ,
\label{va-zeta2}
\end{eqnarray}
where $F\equiv F(a,\zeta ,f_{0},f_{1})$. The condition (\ref{va-zeta2}) leads to
two solutions ($\cos \left( {k\zeta}\right) =\pm 1$), which correspond to changing 
the relative signs of the constants $f_{0}$ and $f_{1}$, and Eqs. (\ref{va-mu}) 
and (\ref{va-N}) can be written as
\begin{eqnarray}
\mu _{\pm } &=&\frac{1}{4a^{2}}+\beta \left[ f_{0}^{2}+\frac{1}{2}
f_{1}^{2}(1+e^{-(ka)^{2}})\pm 2f_{0}f_{1}e^{-(ka/2)^{2}}\right]  \nonumber \\
&+&\left( {\ \sqrt{2\pi }ga}-{2F_{\pm }}\right) \frac{N_{\pm }}{2\pi a^{2}},
\label{va-mu2} \\ \nonumber \\
N_{\pm } &=&2\pi \frac{1+2\beta (ka^{2})^{2}f_{1}(f_{1}e^{-(ka)^{2}}\pm
f_{0}e^{-(ka/2)^{2}})}{4F_{\pm }-2a\partial {F_{\pm }}/\partial {a}-\sqrt{
2\pi }ga},  \label{va-N2}
\end{eqnarray}
where $F_{+}\equiv F(a,0,f_{0},f_{1})$ and $F_{-}\equiv F(a,\pi
/k,f_{0},f_{1})$.

\begin{figure}[tbph]
\centerline{
\includegraphics[width=7.5cm,clip]{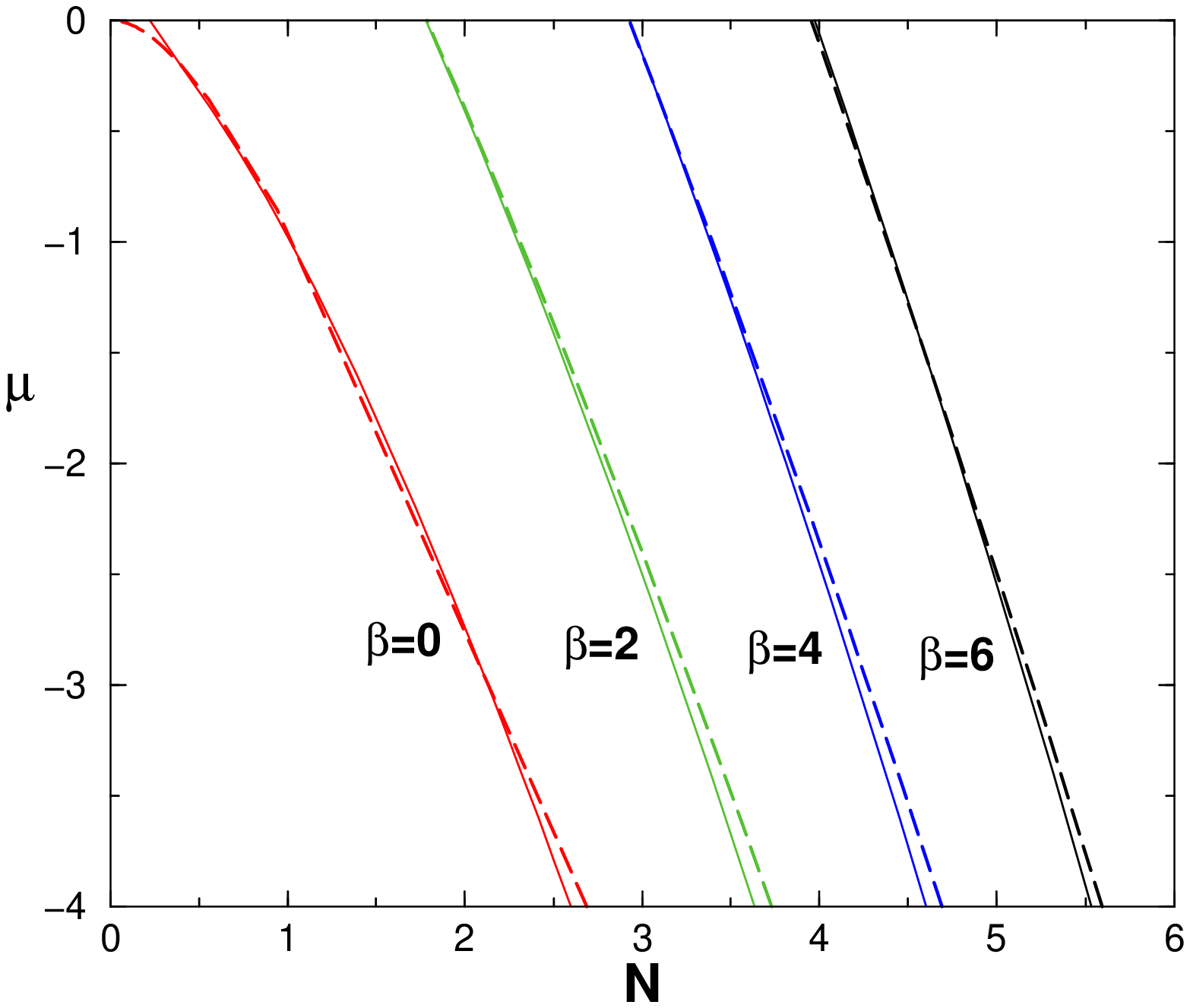}
\includegraphics[width=7.5cm,clip]{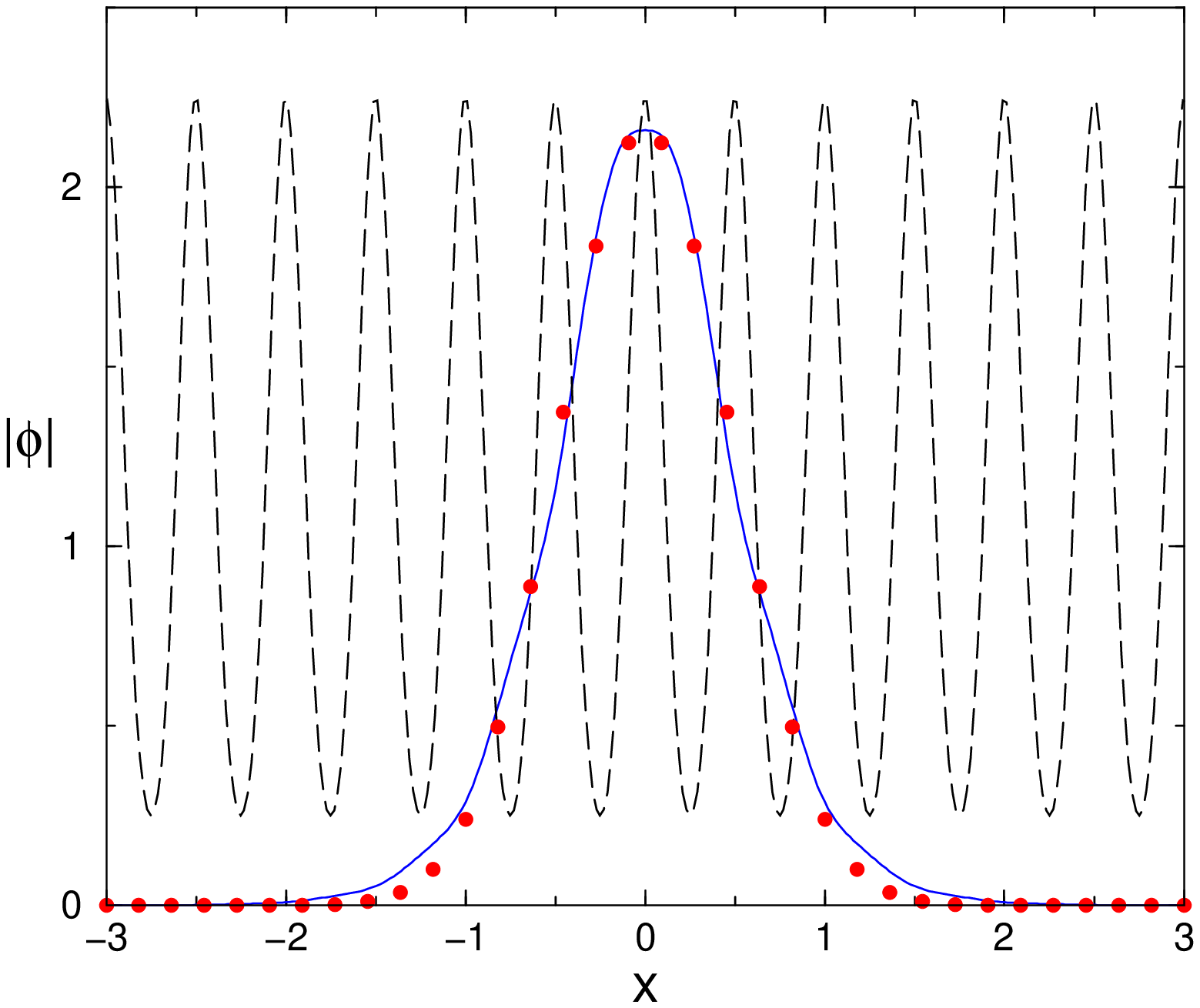}
}
\caption{Left panel: chemical potential $\protect\mu $ as a function of norm
$N$, obtained from direct numerical solutions of Eq. (\protect\ref{bseq})
(solid lines), and from the corresponding variational approach (dashed
lines). This figure stresses the effect of the nonlinearity parameter, $
\protect\beta $ (the scaled polarizability) on the results. Other parameters
are $\protect\alpha =g=0$, $\protect\sigma =1$, $f_{0}=1$, $f_{1}=0.5$ and $
\Lambda =0.5$ (corresponding to $k=4\protect\pi $). Right panel: the
numerical solution (the solid line) for the profile of the wave function,
centered in $x=0$, is compared to the corresponding variational result (the
dotted curve), for $\protect\beta =6$ and $\protect\mu =0$. In this right
panel, we also plot function $f^{2}(x)$ by dashed line. }
\label{fig01}
\end{figure}

\begin{figure}[tbph]
\centerline{
\includegraphics[width=12cm,clip]{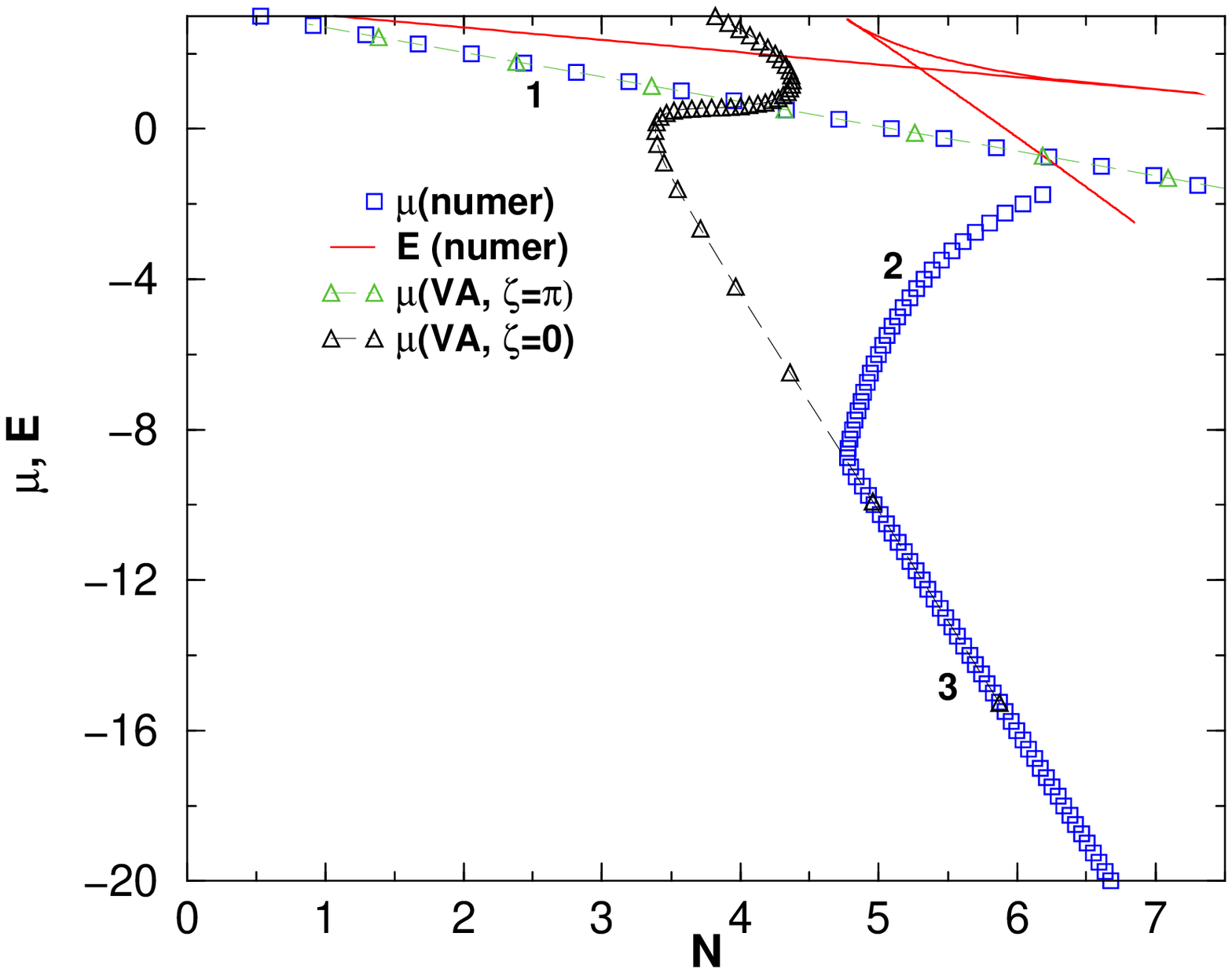}
}
\centerline{
\includegraphics[width=8cm,clip]{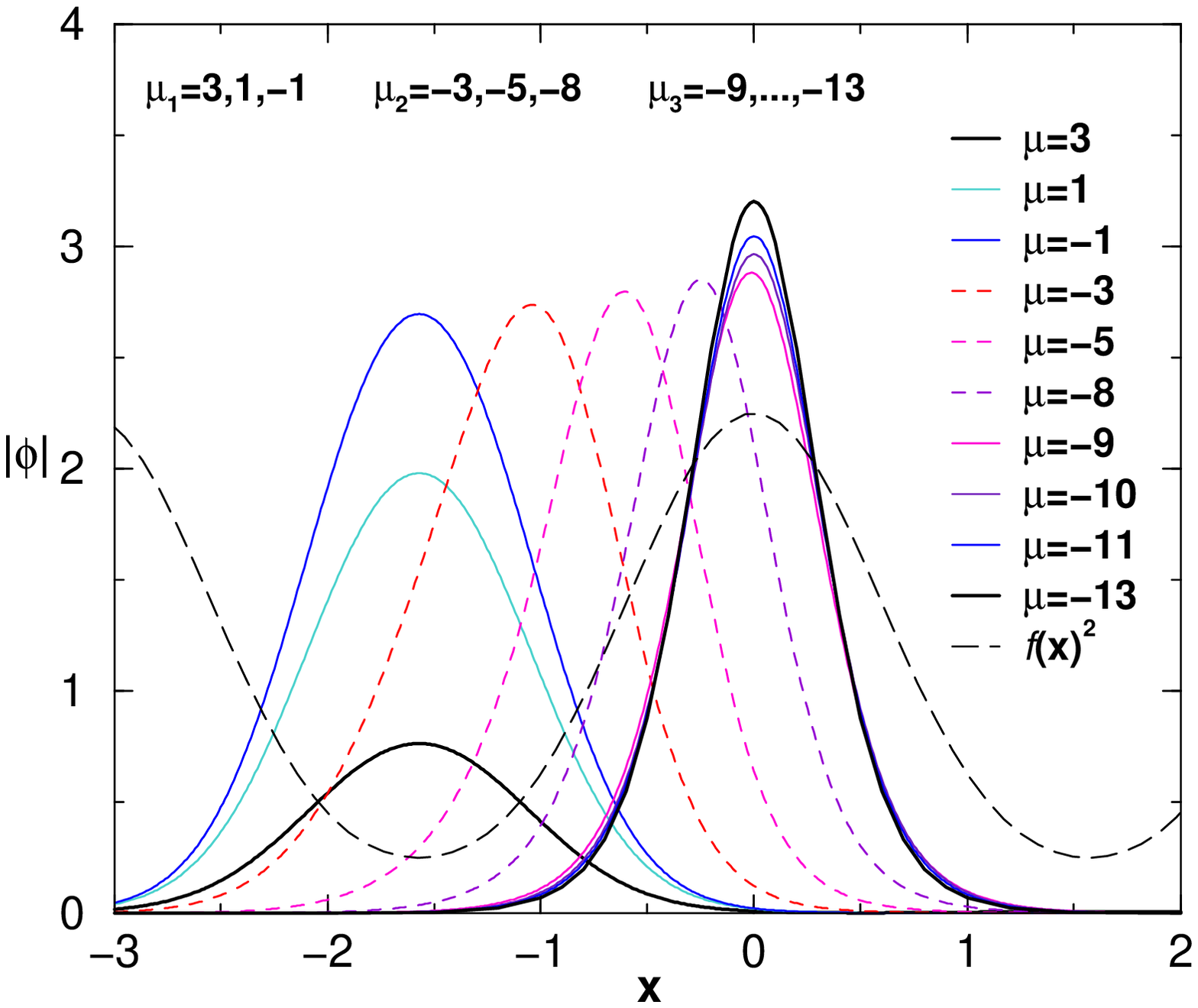}
\hskip 0.1cm
\includegraphics[width=8.5cm,clip]{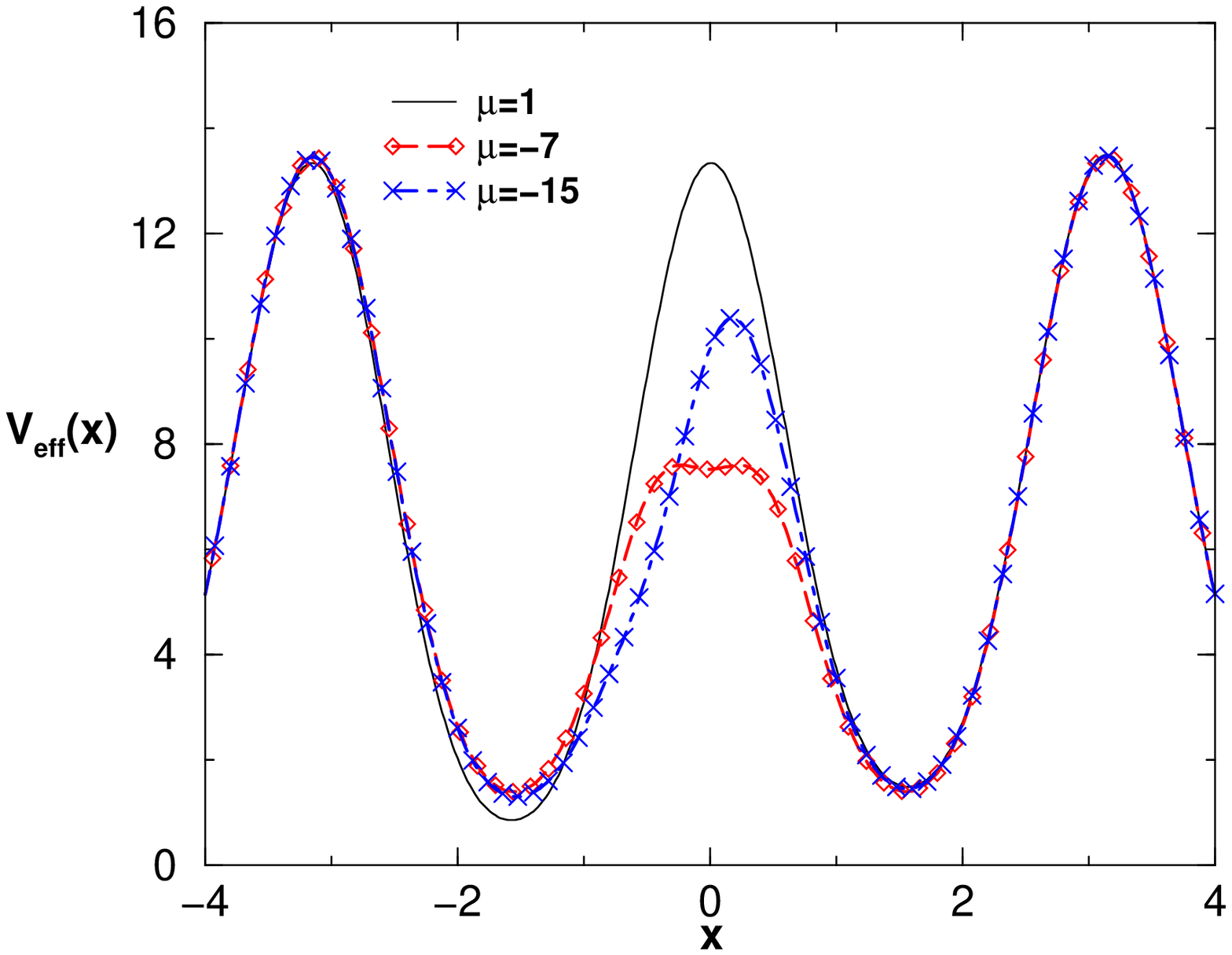}
}
\caption{The top panel: chemical potentials, as obtained from numerical
solutions and from the VA, are shown for $\Lambda =\protect\pi $ (i.e., $k=2$
). Other parameters are $\protect\alpha =g=0$, $\protect\sigma =1$, $f_{0}=1$
, $f_{1}=0.5$ and $\protect\beta =6$. In this panel, we indicate three
regions (1, 2 and 3) for the numerical solutions, following the variation of
the chemical potential ($\protect\mu _{1}$, $\protect\mu _{2}$ and $\protect
\mu _{3}$), to identify the corresponding profiles that are shown in the
left-bottom panel. For the reference, in the top panel we also plot the
total energy (solid-red line), as obtained from the numerical solution. The
modulation function, $f^{2}(x)$, is shown by the black-dashed line in the
left bottom panel. The numerical results for the corresponding effective DDI
potential, given in Eq.(\protect\ref{ddipot}), are shown in the right-bottom
panel for three different values of the chemical potential. }
\label{fig02}
\end{figure}

The solutions produced by the VA are compared with their numerically found
counterparts in Figs.~\ref{fig01} and \ref{fig02}, for $g=\alpha =0$, $
f_{0}=1$, $f_{1}=0.5$, and $\Lambda =0.5$ ($k=4\pi $). The corresponding
numerical solutions of Eq.~(\ref{bseq}) were obtained by means of the
relaxation technique, as described in Ref.~\cite{brtka}. The effect of
parameter $\beta $ is illustrated by plots of chemical potential $\mu $
versus the number of atoms, $N$, in the left panel of Fig.~\ref{fig01}. The
figure demonstrates that the VA provides good agreement with the numerical
results. Variational and numerically found profiles of the solitons are
compared in the right panel of the figures, for $\mu =0$ and $\beta =6$. In
the right panel, we also show the oscillatory function, $f^{2}(x)$, by the
dashed line.

The variational approach produces more accurate results for large
$\beta $ because, in this case, the contribution of the linear
lattice grows, and it is known that the Gaussian ansatz, that we use
here, works well with linear lattice potentials \cite{KMT}. Further,
the steady increase of $\mu $ with $\beta $ in the left panel of
Fig. \ref{fig01} is also explained by the fact that the linear
potential in Eq. (\ref{eq1}) is multiplied by $\beta $.

In the top panel of Fig.~\ref{fig02}, for $k=2$ $(\Lambda =\pi )$ and $\beta
=6$, with the other parameters the same as in Fig.~\ref{fig01}, the
numerical results for the $\mu (N)$ dependences are shown in three distinct
regions: two stable (1 and 3) and one unstable (region 2). The profiles,
displayed in the left bottom panel, clearly show that the wave function
profiles are centered at $x=\pi /2$ in stable region 1. In unstable region
2, we verify a transition from stable region 1 to stable region 3, with the
center of the profiles moving to $x=0$. For the reference's sake, the
effective DDI potential, as defined by Eq.~(\ref{ddipot}), is displayed in
the right bottom panel for three values of the chemical potential. A
noteworthy feature is asymmetry with respect to the reflection, 
$x\rightarrow - x$, which can also be observed in the profiles shown in the
left bottom panel of Fig.~\ref{fig02}.

As seen in Fig.~\ref{fig02}, the VA gives a perfect agreement with the
numerical solutions in region 1 (where the center of the profile is located
at $x=\zeta =\pi /2$ ), when $\mu >-2$, and in region 3 (where the center of
the profile is at $x=\zeta =0$), when $\mu <-9$. However, the VA cannot
follow the behavior presented by the numerical results in region 2, because
the simple Gaussian ansatz is not an adequate one, in this case.\textbf{\ }
In the top panel, we also present the corresponding total energy, obtained
from the numerical solutions. In the left bottom panel, together with the
profiles, $f^{2}(x)$ is displayed by the dashed line.

\begin{figure}[tbph]
\centerline{
\includegraphics[width=12cm,clip]{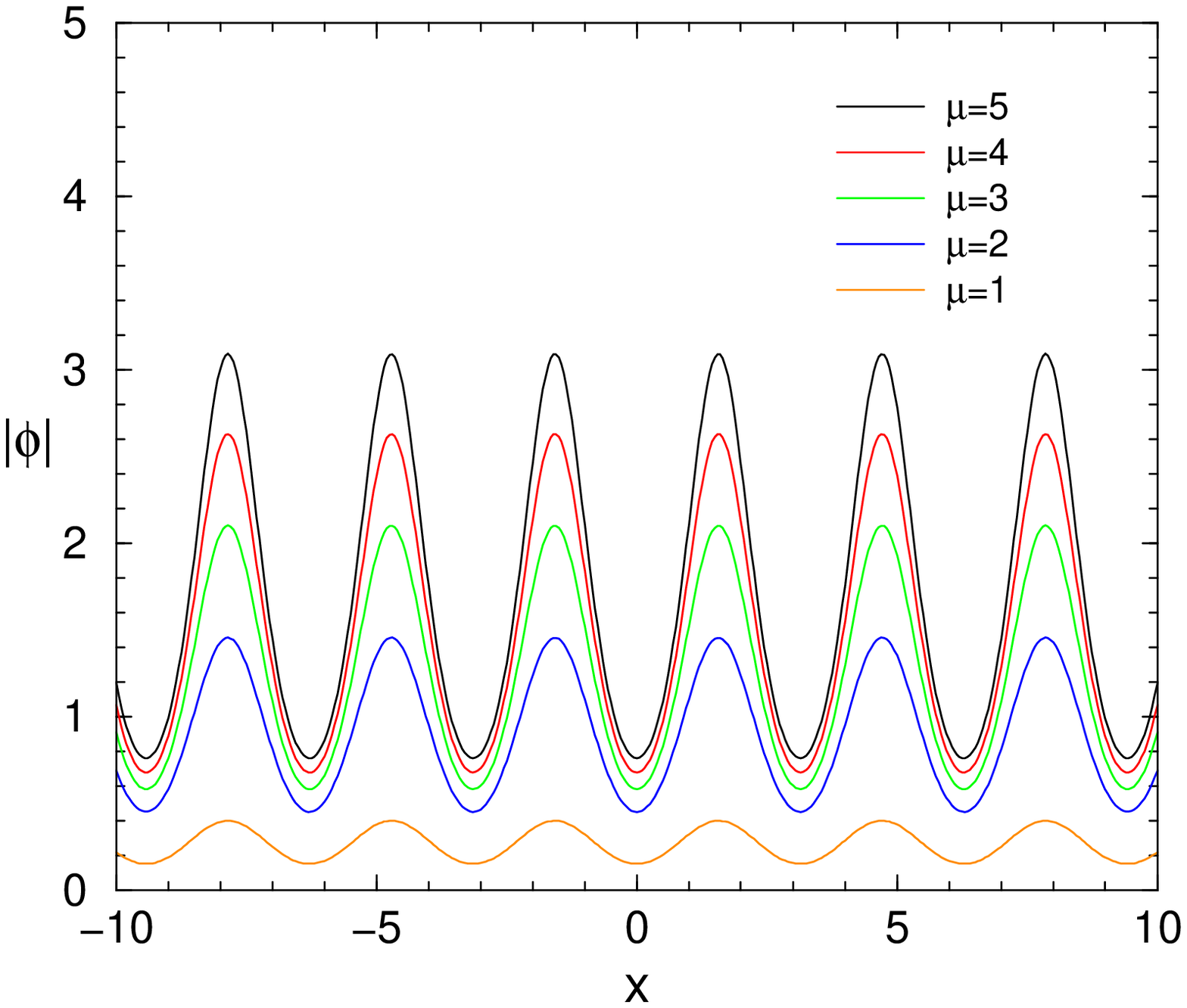}
}
\caption{(Color online) Solution profiles for $\protect\sigma =-1/2$, with
other parameters the same as in Fig.~\protect\ref{fig02} (in particular, 
$\Lambda =\protect\pi $ and $\protect\beta =6$). }
\label{fig03}
\end{figure}

For the perpendicular orientation of the dipoles, corresponding to the
repulsive DDI, $\sigma =-1/2$ (while other parameters are the same as in 
\ref{fig02}), we demonstrate in Fig.~\ref{fig03}, by means of numerical results
for different values of $\mu $, that the wave-function profiles are
delocalized, i.e., they do not build bright solitons.
\begin{figure}[tbph]
\centerline{
\includegraphics[width=8cm,clip]{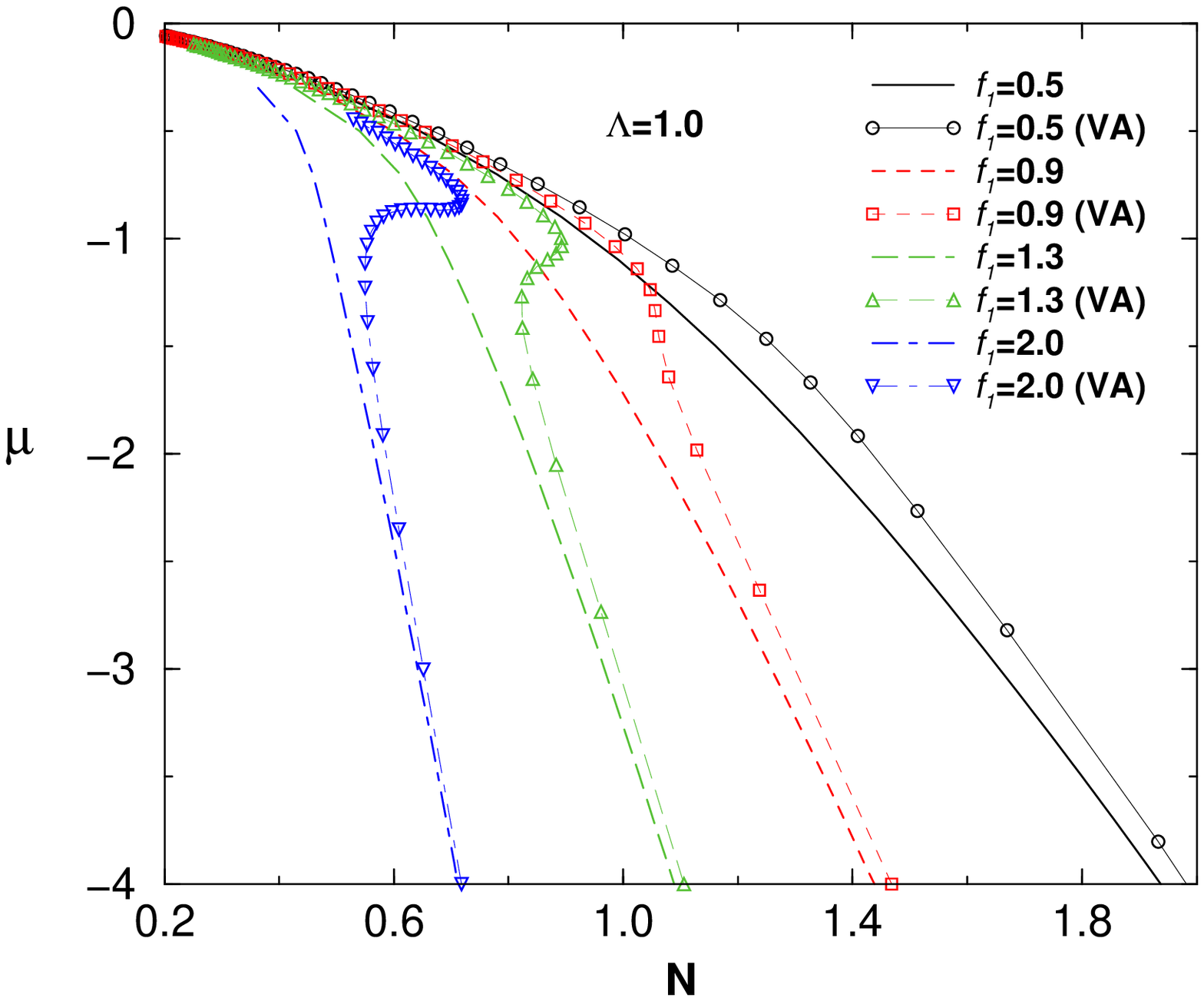}
\includegraphics[width=8cm,clip]{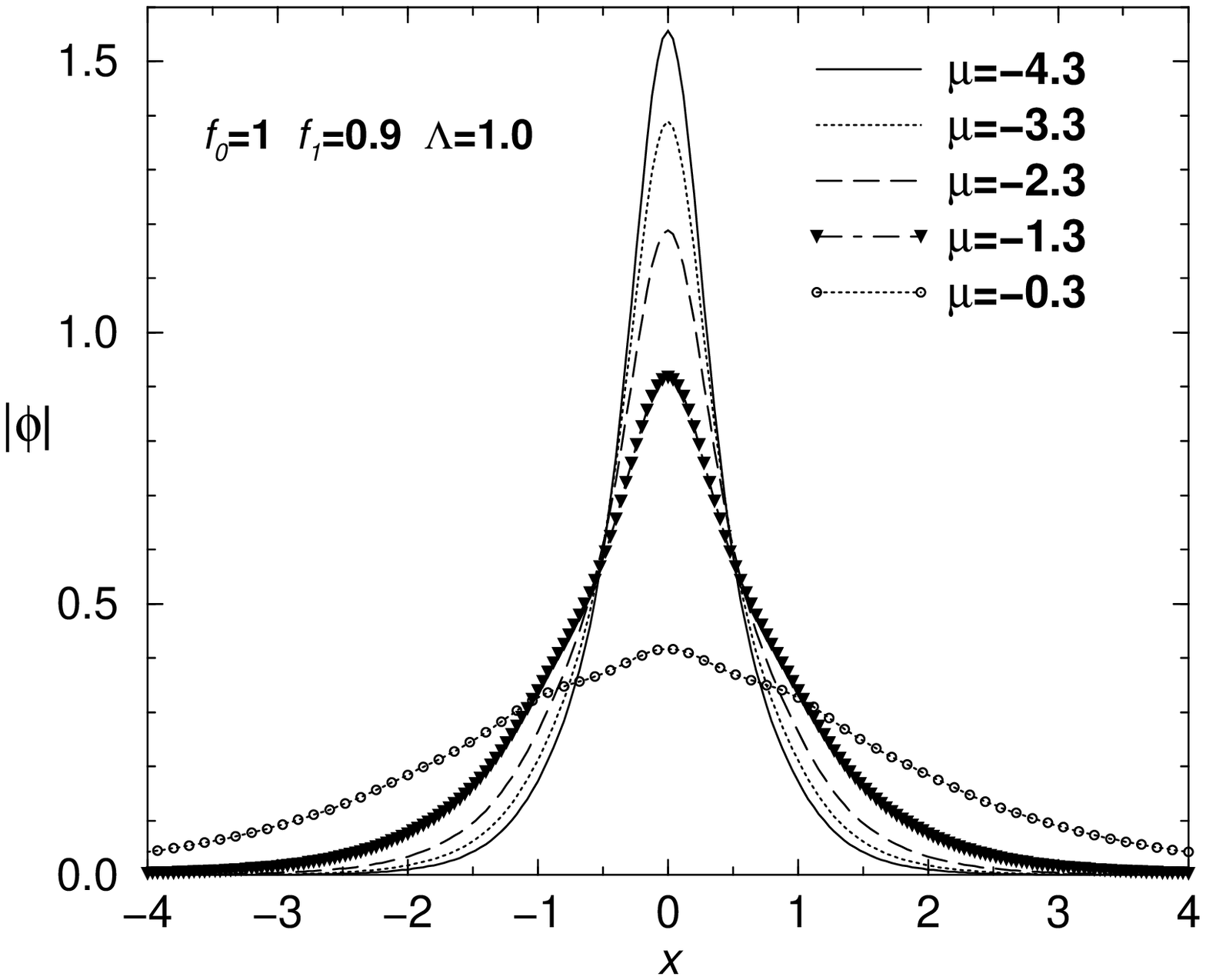}
}
\centerline{
\includegraphics[width=8cm,clip]{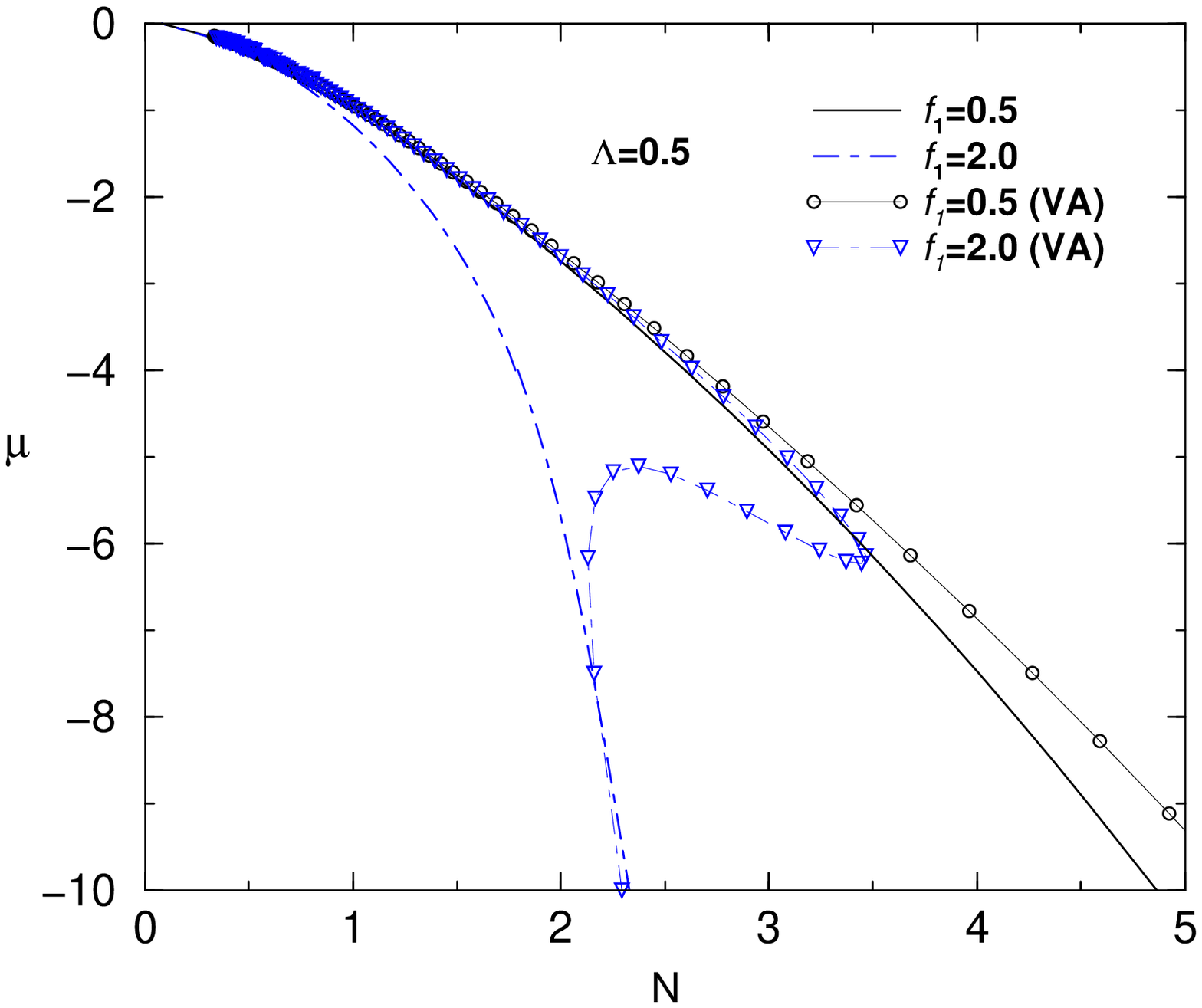}
\includegraphics[width=8cm,clip]{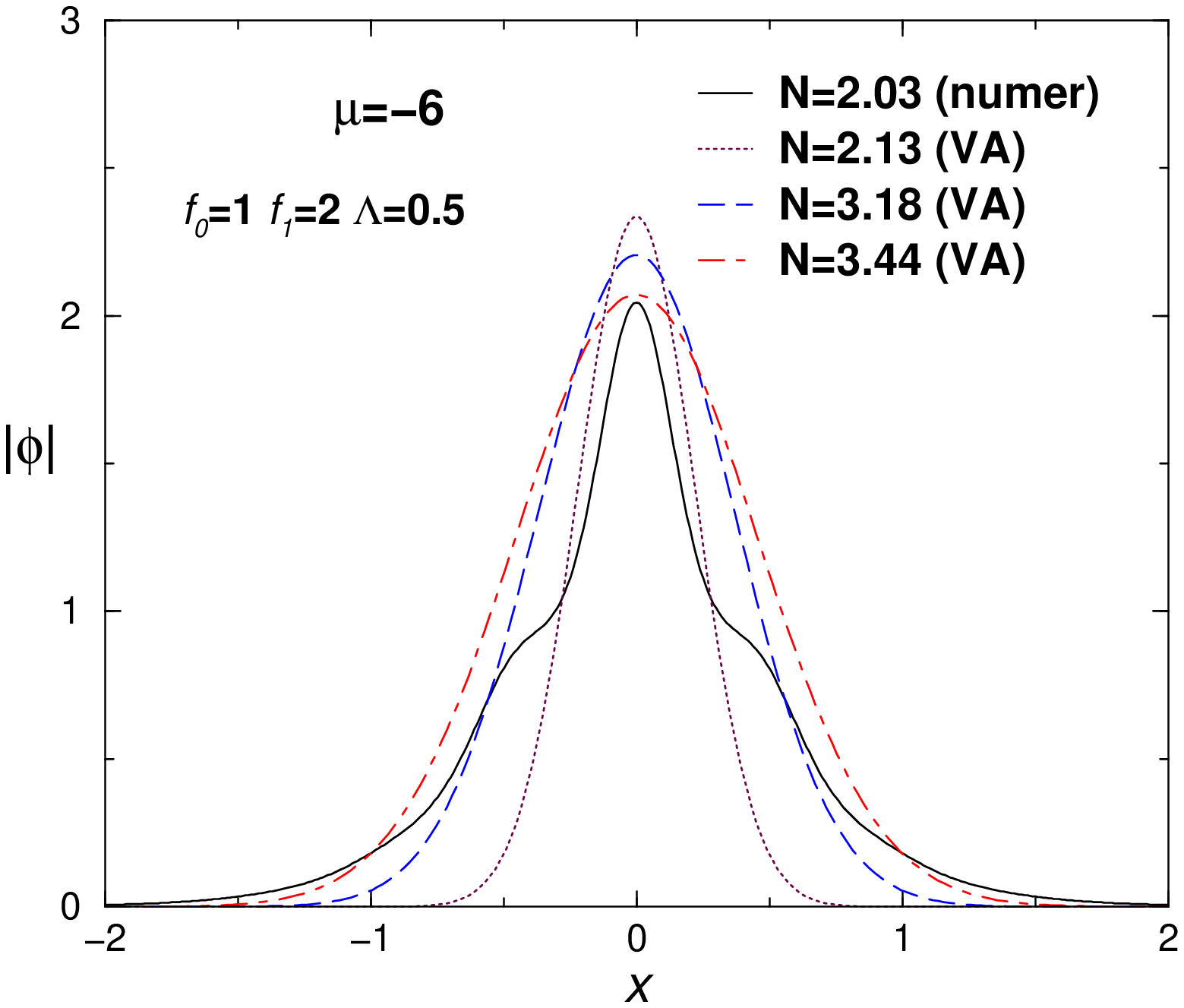}
}
\caption{(Color online) Left panels: plots $\protect\mu (N)$, for different
values of $f_{1}$ (as indicated in the frames), with $\protect\beta =0$, 
$f_{0}=1$, and $\Lambda =1$ or $0.5$ (the top and bottom panels,
respectively). In the right top frame, numerically found profiles are
displayed for different values of $\protect\mu $, corresponding to the case
with $f_{1}=0.9$, which is shown in the left-top panel. In the right-bottom
frame, we consider the case with $\protect\mu =-6$ and $f_{1}=2$, where
three variational profiles are displayed for different values of $N$. }
\label{fig04}
\end{figure}

In Fig.~\ref{fig04}, we present numerical results for $\mu (N)$ (the left
panels), compared with VA findings for $\beta =0$, in the absence of the
linear trap ($\alpha =0$). In the left panels we consider different values
of $f_{1}$, with $\Lambda =1$ (left top) and $0.5$ (left bottom). According
to the Vakhitov-Kolokolov (VK) criterion, the solutions may be stable if the
condition ${\partial \mu }/{\partial N}<0$ holds \cite{vakhitov}. This
assumption is well verified by our numerical results, for the whole range of
parameters that we have analyzed. Simulations of the corresponding temporal
evolution (not shown in this figure) validate the VK criterion in the
present model. However, the VA results cannot follow the results to the full
extension, besides the fact that they present very good agreement for large
negative values of $\mu $. As seen in the left top panel of Fig.~\ref{fig04}
, for $\Lambda =1$ the VA correctly predicts the stability and converges to
numerical results for $\mu <-1.5$ at all values of $f_{1}$. In the case of 
$\Lambda =0.5$ (the left bottom panel), the VA results are equally accurate
at $\mu <-6$. On the other hand, in the case of $f_{1}=2.0$ and $\Lambda
=0.5 $, the VA results represent a set of two solutions, one being nearly
insensitive to variations of $f_{1}$.

\subsection{The pure nonlocal nonlinear lattice ($\protect\alpha = \protect
\beta =0$)}

\begin{figure}[tbph]
\centerline{
\includegraphics[width=9cm,clip]{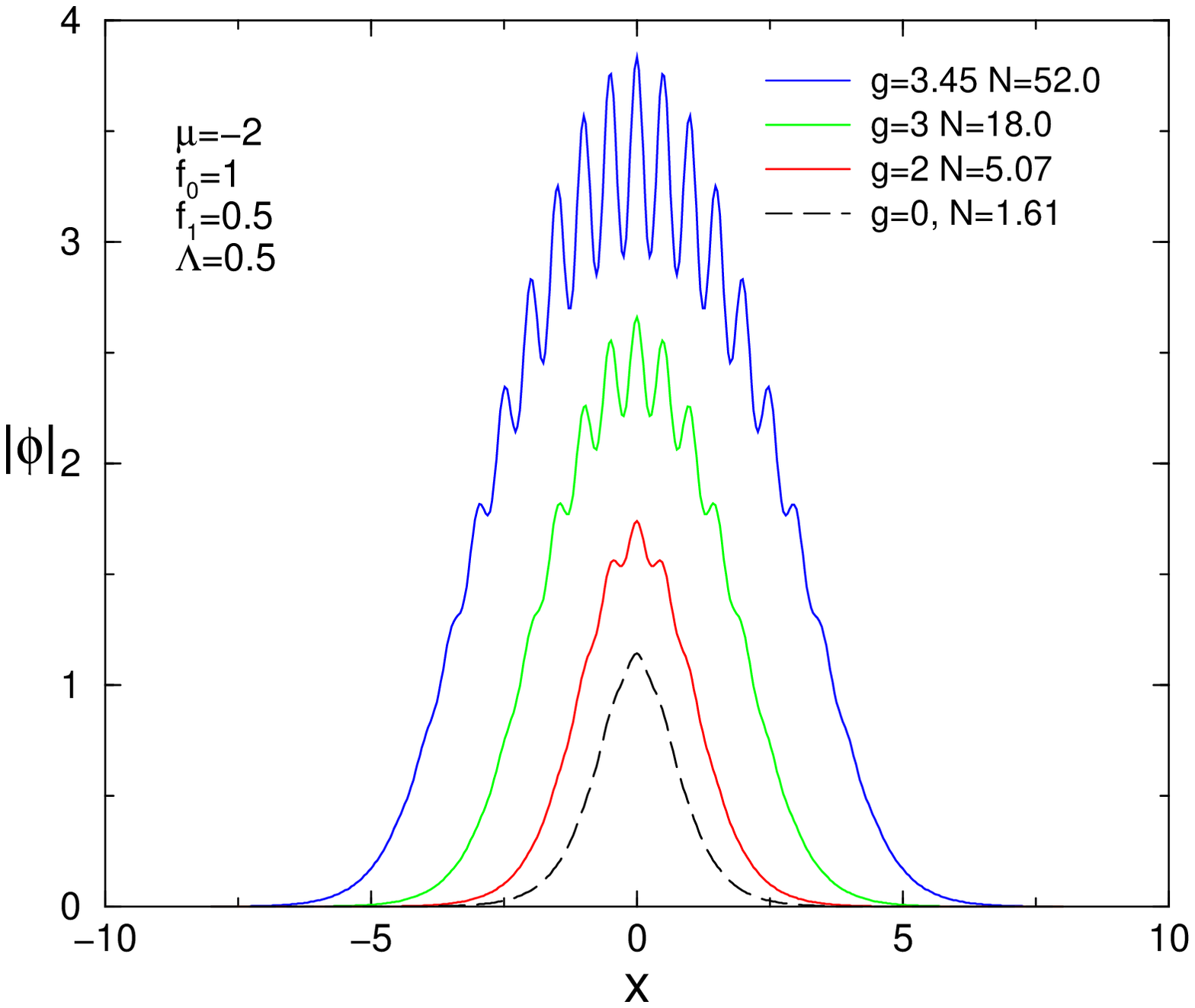}
\includegraphics[width=8.5cm,clip]{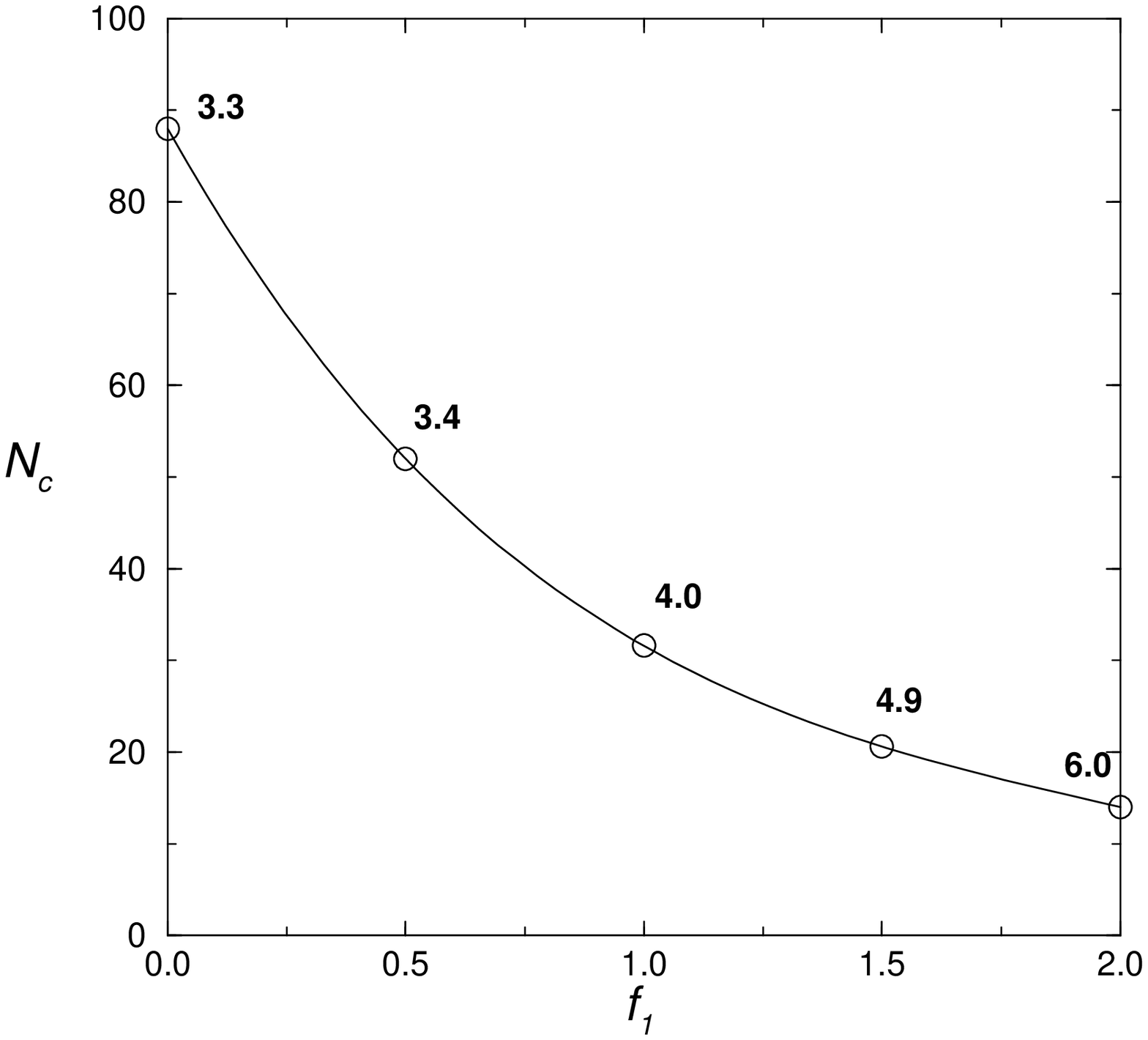}
}
\caption{(Color online) The left panel: numerical results for the
wave-function profiles with fixed $\protect\mu =-2$ and a few values of the
contact-interaction strength $g$. The corresponding values of $N$ are
indicated inside the panel. Other parameters are $f_{0}=1$, $f_{1}=0.5$ and $
\Lambda =0.5$. These results for $f_{1}=0.5$ define critical maximum values $
g_{c}=3.45$ and $N_{c}=52$, above which no bright soliton were found. In the
right panel, varying modulation amplitude $f_{1}$ in the interval of $0\leq
f_{1}\leq 2$ and keeping other parameters as in the left panel, we display a
curve corresponding to the critical values of $N$ and $g$. Critical values $
g_{c}$ are indicated for some data points along the curve. }
\label{fig05}
\end{figure}

Fixing $\alpha =0$ and $\beta =0$, we have analyzed the model with the
repulsive contact interactions ($g>0$). The corresponding term in Eq.~(\ref
{eq3}) tends to expand the wave function, on the contrary to the attractive
nonlocal nonlinear interaction, cf. Ref. \cite{Cuevas}. In Fig.~\ref{fig05},
fixing other parameters as $f_{0}=1$, $f_{1}=0.5$, $\Lambda =0.5$, and $\mu
=-2$, we present stationary solutions for different magnitudes of $g$, in
the left panel. As $g$ increases, the wave function indeed gets broader and
the number of atoms trapped in the soliton increases. Beyond a critical
value, $g_{c}=3.45$, no solution can be found. In the right panel of 
Fig.~\ref{fig05}, keeping the chemical potential $\mu $ and the other parameters,
given in the left panel, fixed, we present $g_{c}$ and the corresponding
value of $N$ for different values of $f_{1}$. It is observed that $N$
decreases and $g_{c}$ increases with the increase of $f_{1}$.

This dependence can be explained considering the broad soliton case. Then
the nonlocal term can be approximated as the local one with an effective
nonlinearity coefficient,
\[
\gamma _{\mathrm{eff}}=f^{2}(x)\int_{-\infty }^{+\infty }dyR(y)
\]
and the bright solitons should exist, provided that $g>f^{2}\int_{-\infty
}^{+\infty }dyR(y)$. This arguments explain the growth of $g_{c}$ with the
increase of $f_{1}$, as observed in the right panel of Fig. \ref{fig05}.

\section{Dynamics of bright solitons}

In this section we address the mobility of solitons and their collisions.
The soliton motion in nonlinear lattices was previously considered in Refs.
\cite{SM,AG,AGBT,KT10}. First, we present full numerical solutions of the 1D
GPE (\ref{eq1}), exploring a parameter region for finding stable
bright-solitons solutions. This is followed by consideration of a dynamical
version of the variational approach, with some results for frequencies of
oscillations of perturbed solitons being compared with the full numerical
calculations.

\begin{figure}[tbph]
\centerline{
\includegraphics[width=8cm,clip]{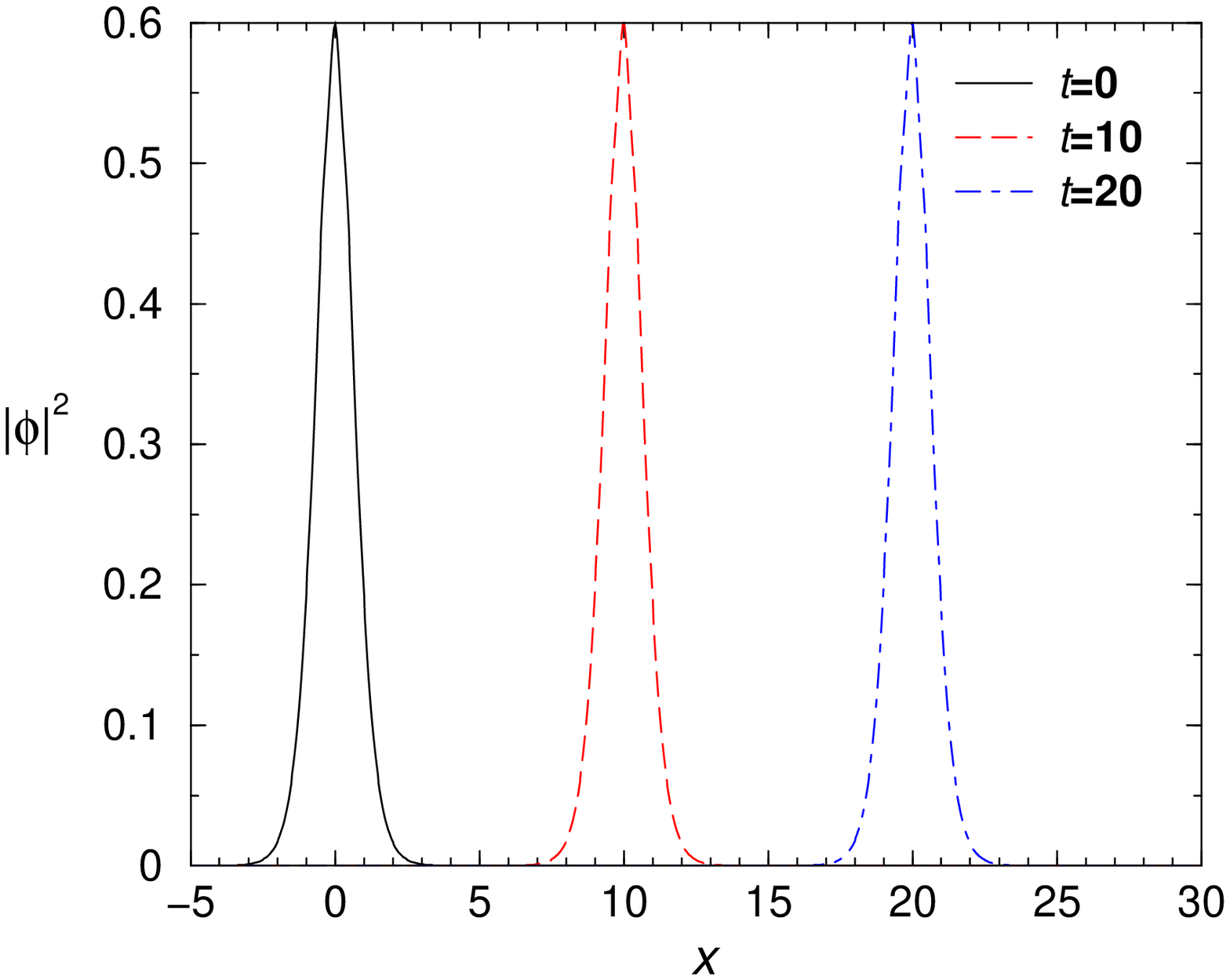}
\includegraphics[width=8cm,clip]{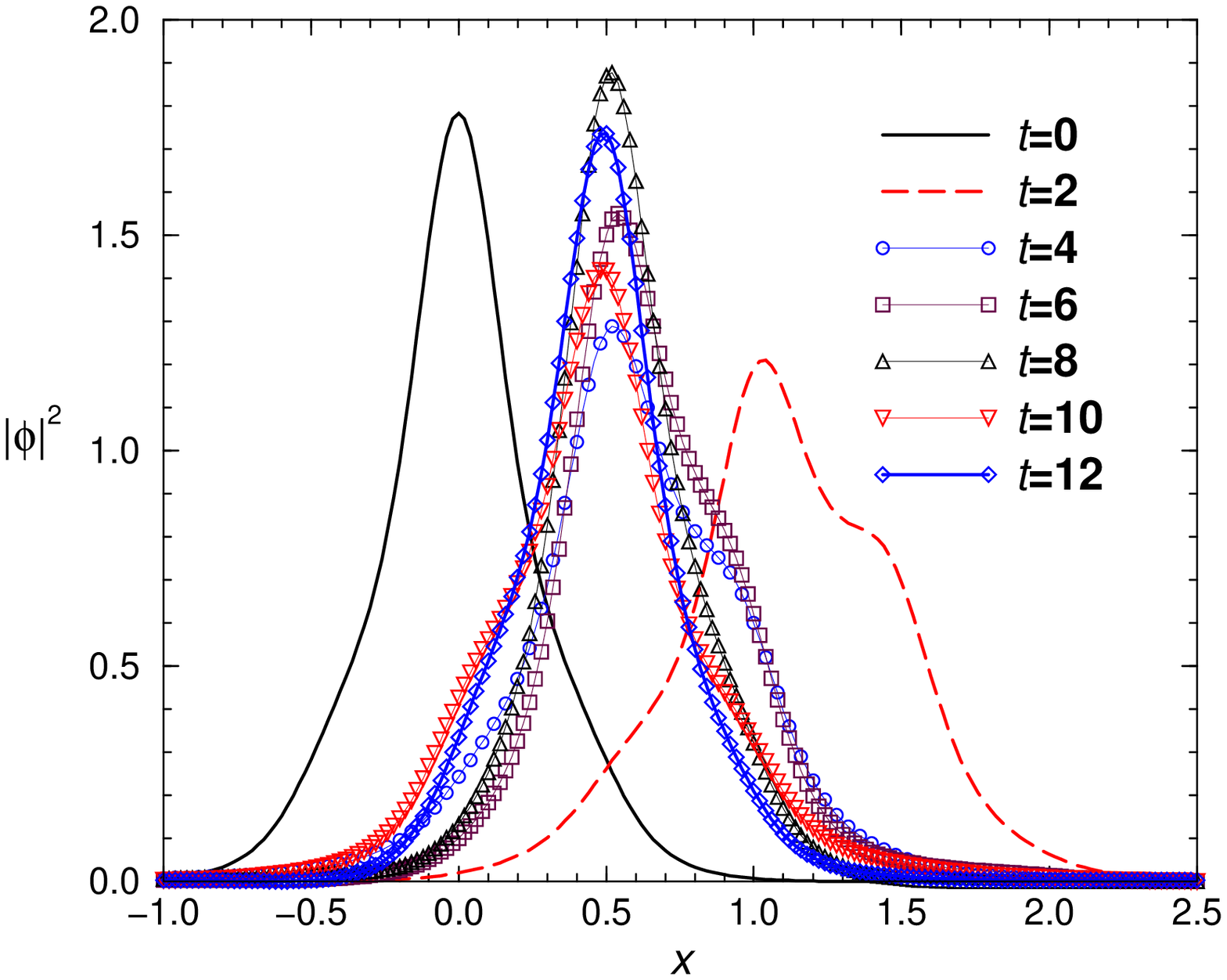}
}
\caption{(Color online) Propagation of a soliton. In the left panel, soliton
profiles are shown with time intervals $\Delta t=10$ (the velocity is 1),
for $\protect\mu =-1$ and $N\approx 1.02$. In the right panel, $\protect\mu 
=-10$, $N\approx 4.86$, with the profiles shown with intervals $\Delta t=2$.
In both cases, $f_{0}=1$, $f_{1}=0.5$ and $\Lambda =0.5$ (with $\protect
\alpha =\protect\beta =0$). }
\label{fig06}
\end{figure}

The propagation of a soliton is presented in two panels of Fig.~\ref{fig06}.
In the left panel, for $\mu =-1$ and $N\approx 1.02$, we show the soliton
propagation by considering, in the dimensionless units, a time interval from
$0$ to $20$, and velocity equal to $1$. In the right panel, we present the
case of $\mu =-10$ and $N\approx 4.86$, showing profiles separated by time
intervals $\Delta t=2$. In the latter case, the soliton ends up getting
trapped at a fixed position. In both the cases, we have $\alpha =\beta =g=0$
, and $f_{0}=1$, $f_{1}=0.5$, $\Lambda =0.5$.

\begin{figure}[tbph]
\centerline{
\includegraphics[width=8cm,clip]{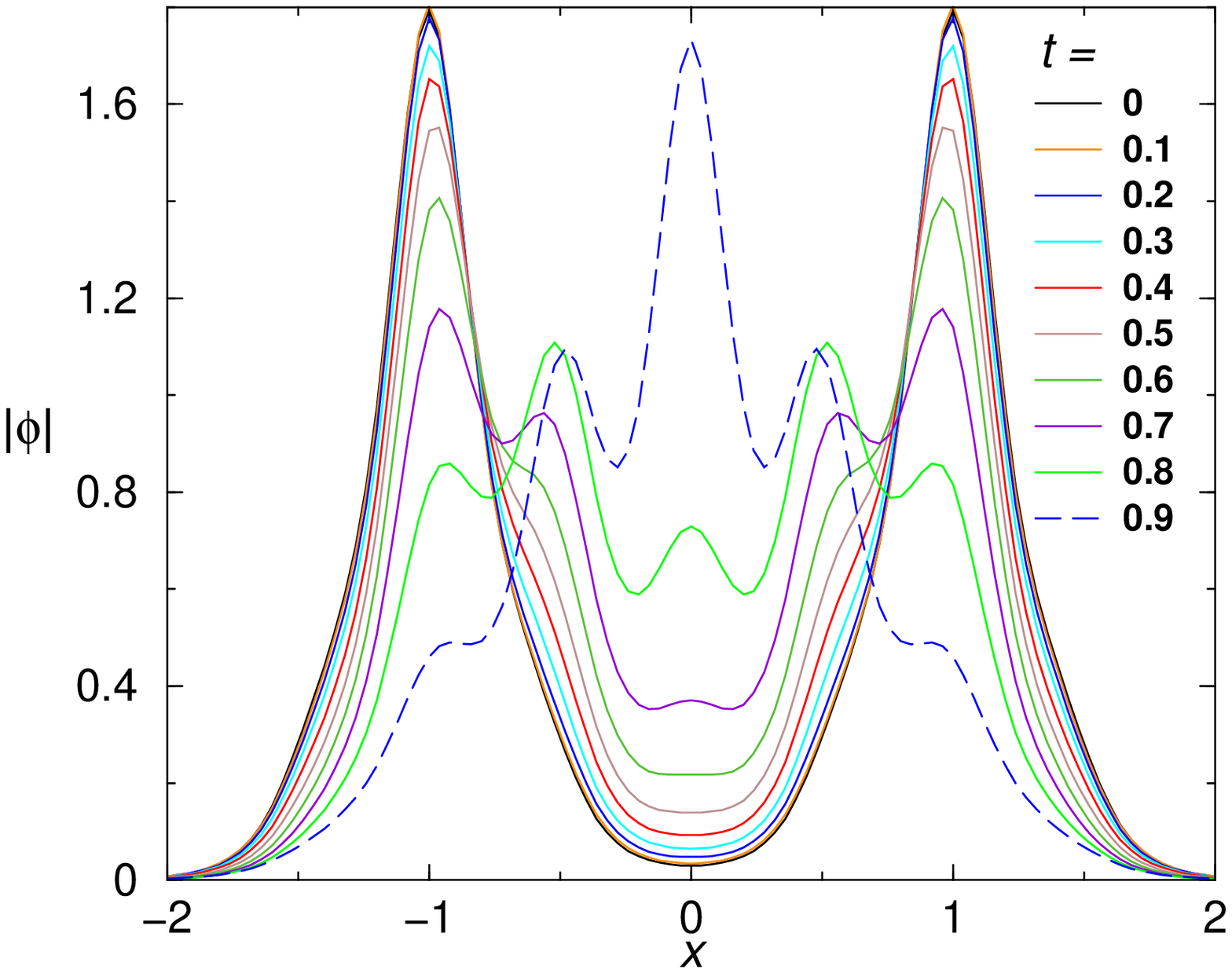}
\includegraphics[width=7.7cm,clip]{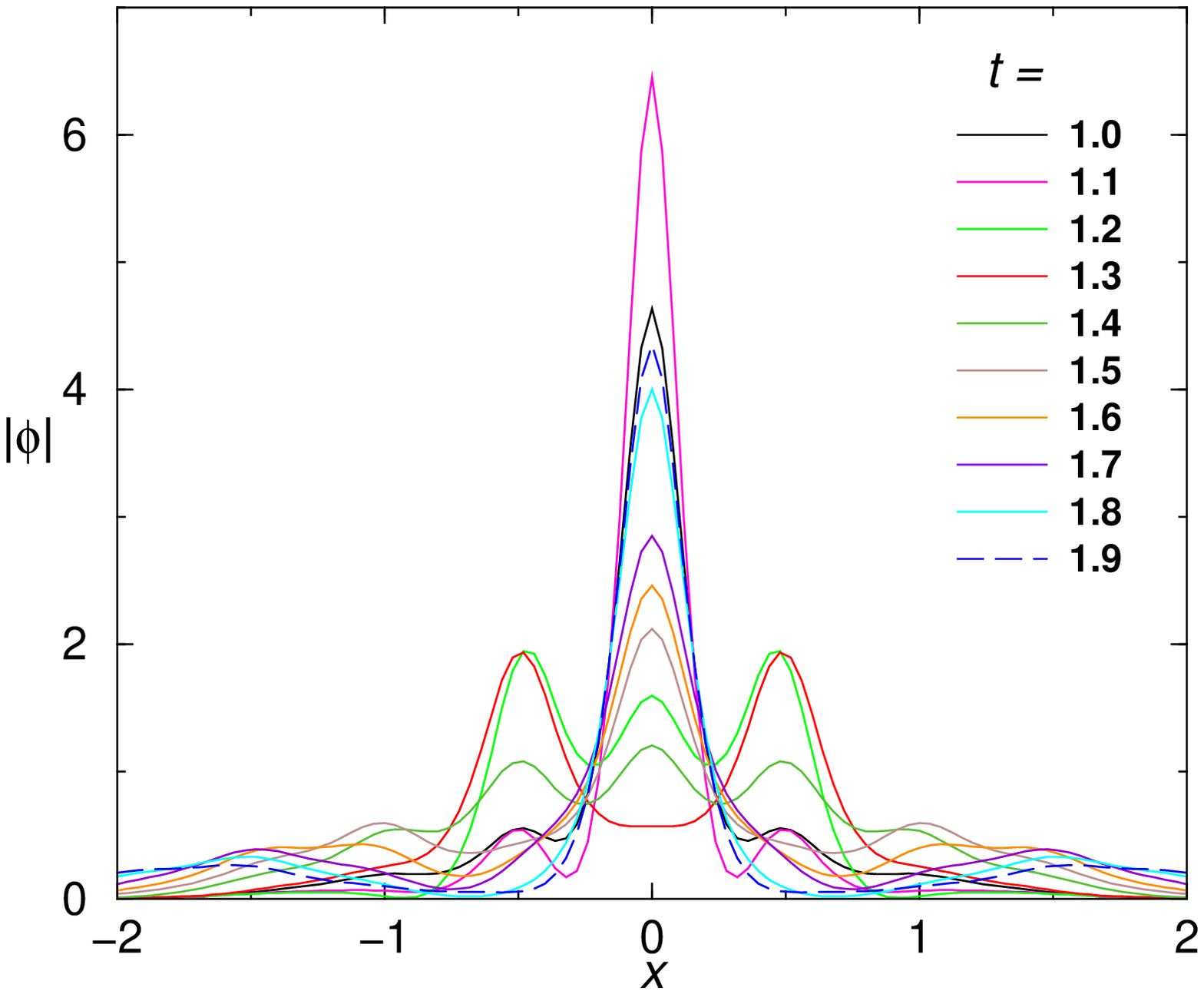}
}
\centerline{
\includegraphics[width=8cm,clip]{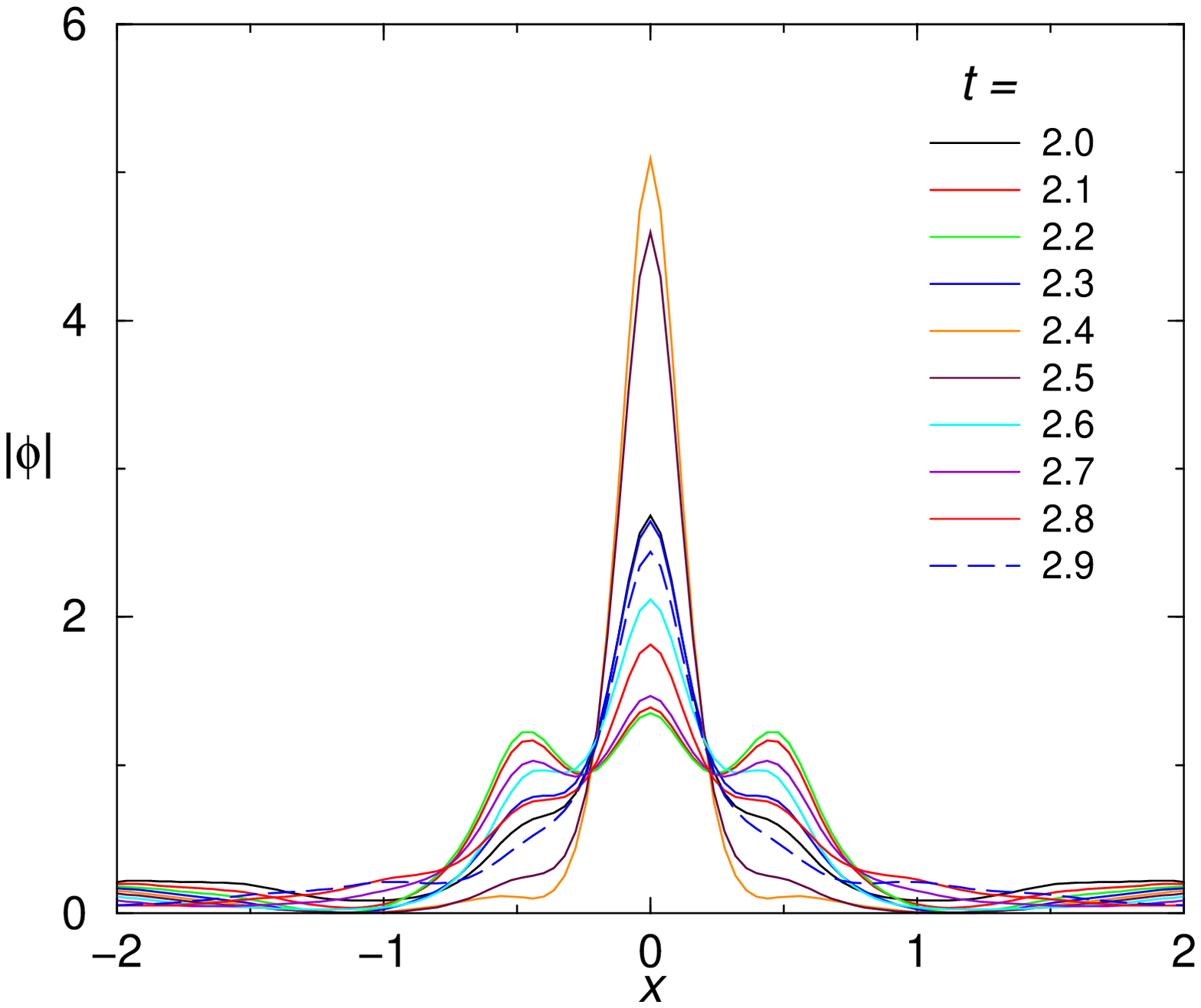}
\includegraphics[width=8cm,clip]{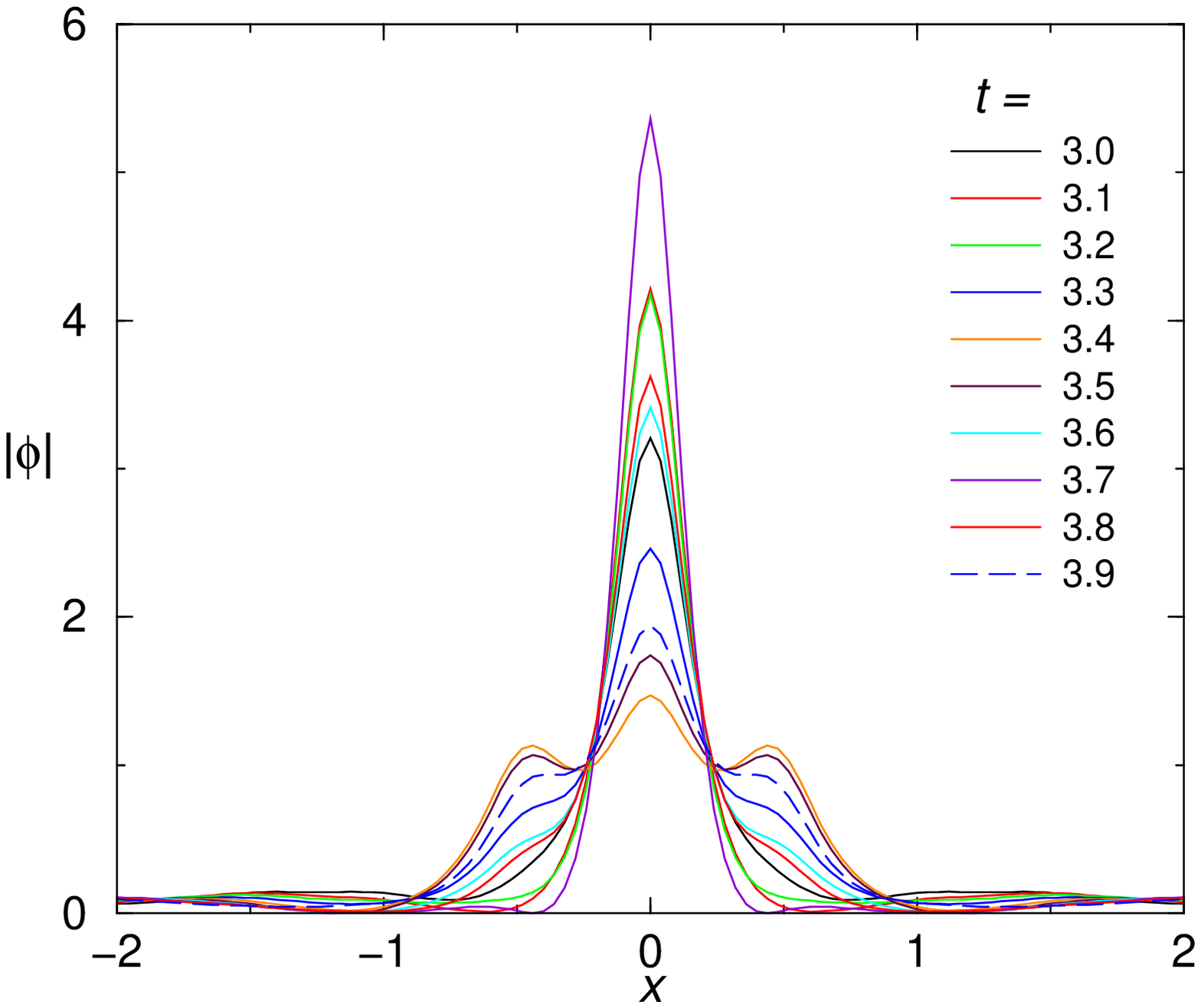}
}
\caption{(Color online) The interaction of two solitons, with $\protect\mu
=-10$, $N\approx 4.86$, is shown in four panels at different moments of
time, with $\Delta t=0.1$ (the average velocity is zero), as indicated in
the panels. In all the cases, parameters are $f_{0}=1$, $f_{1}=0.5$ and $
\Lambda =0.5$ ($a_{\perp }=2,\protect\lambda =1$). The phase difference
between the solitons is zero, therefore they attract each other. One can see
a transition from a bound state to a breather. }
\label{fig07}
\end{figure}

\begin{figure}[tbph]
\centerline{
\includegraphics[width=16cm,clip]{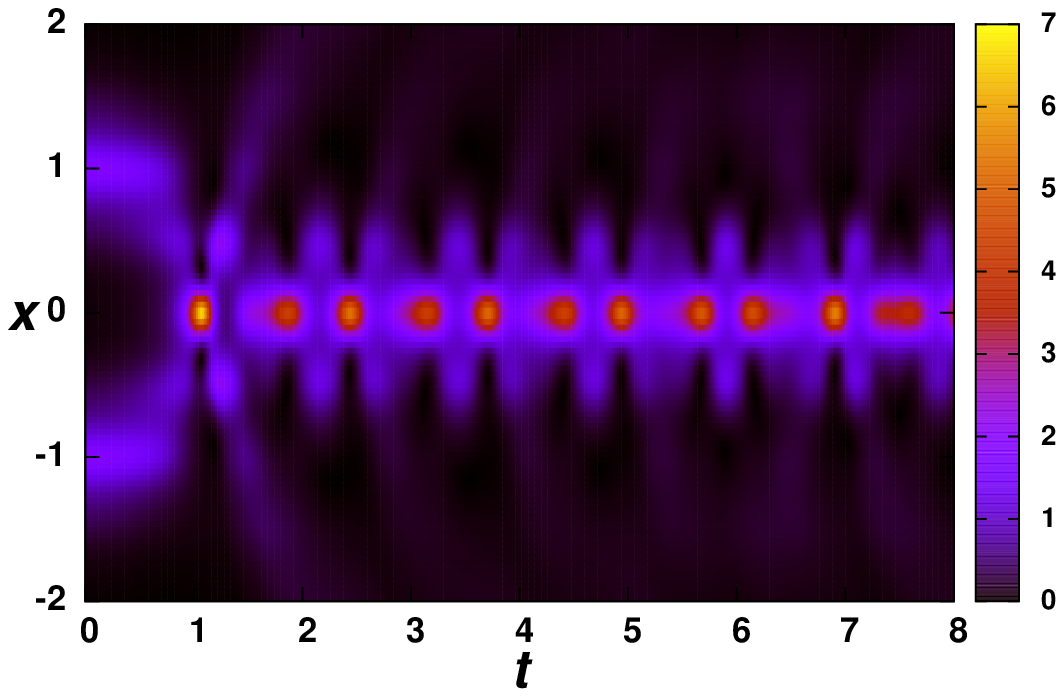}
} \vspace{-1cm}
\caption{(Color on-line) The density plot corresponding to the results shown
in Fig.~\protect\ref{fig07}. }
\label{fig08}
\end{figure}

The interaction of two solitons is shown in four panels of Fig.~\ref{fig07},
for $\mu =-10$ and $N\approx 4.86$. The field profiles are displayed with
time intervals $\Delta t=0.1$, the average velocity being zero. The
parameters are the same as in Fig.~\ref{fig06} ($f_{0}=1$, $f_{1}=0.5$ and $
\Lambda =0.5$). We consider the solitons with zero phase difference between
them, hence they attract each other. These panels display a transition from
a bound state to a breather. The density plot corresponding to the results
presented in Fig.~\ref{fig07} is displayed in Fig.~\ref{fig08}. For the same
parameters, we have verified that the solitons demonstrate almost no
interaction when the phase shift between them is $\pi $.

Next, for the comparison with numerical solutions, we here present a
dynamical version of the VA, which is based on the following Gaussian
ansatz:
\begin{equation}
\hspace{-1cm}\phi (x,t)=A(t)\exp \left\{ -\frac{\left[ x-\zeta (t)\right]
^{2}}{2\left( a(t)\right) ^{2}}+\mathrm{i}b(t)\left[ x-\zeta (t)\right] ^{2}+
\mathrm{i}\kappa (t)\left[ x-\zeta (t)\right] -\mathrm{i}\mu t\right\} .
\label{ans}
\end{equation}
To derive evolution equations for soliton parameters $A,a,\zeta ,b,k,p$, we
calculate the respective averaged Lagrangian, taken with the full dynamical
density
\begin{eqnarray*}
\mathcal{L}(x,t) &=&\mathrm{Im}\left( \phi \frac{\partial \phi ^{\ast }}{
\partial t}\right) -\frac{1}{2}\left\vert \frac{\partial {\phi }}{\partial x}
\right\vert ^{2}-(\alpha x^{2}+\beta f(x)^{2})|\phi |^{2}-\frac{g}{2}|\phi
|^{4} \\
&+&\frac{f(x)|\phi |^{2}}{2}\int_{-\infty }^{\infty }dx^{\prime
}R(x-x^{\prime })f(x^{\prime })|\phi (x^{\prime })|^{2},
\end{eqnarray*}
cf. its static counterpart (\ref{density}). The averaged Lagrangian per
particle is given by
\begin{eqnarray}
\frac{L}{N} &=&-\frac{1}{2}a^{2}\frac{db}{dt}+\kappa (t)\frac{d\zeta }{dt}
+\mu _{r}-\frac{1}{4a^{2}}-a^{2}b^{2}-\frac{1}{2}\kappa (t)^{2}-\frac{\alpha
}{2}a^{2}-\alpha \zeta ^{2}  \nonumber \\
&-&2\beta f_{0}f_{1}\cos (k\zeta )e^{-a^{2}k^{2}/4}-\frac{\beta }{2}
f_{1}^{2}\cos (2k\zeta )e^{-a^{2}k^{2}}-\frac{gN}{2\sqrt{2\pi }a}+\frac{NF}{
2\pi a^{2}},  \label{dynam}
\end{eqnarray}
where, as in the static setting, we have $N=\sqrt{\pi }A^{2}a$, with $
F\equiv F(a,\zeta ,f_{0},f_{1})$ given by Eq.~(\ref{F}). The Euler-Lagrange
equations following from Lagrangian (\ref{dynam}) give rise to the coupled
evolution equations for the soliton's width and location of the center of
mass:

\begin{eqnarray}
a_{tt}\equiv \frac{d^{2}a}{dt^{2}} &=&\frac{1}{a^{3}}-2\alpha a+2ak^{2}\beta
f_{1}\left[ f_{0}\cos (k\zeta )e^{-a^{2}k^{2}/4}+f_{1}\cos (2k\zeta
)e^{-a^{2}k^{2}}\right]  \nonumber \\
&+&\frac{gN}{\sqrt{2\pi }a^{2}}-\frac{2NF}{\pi a^{3}}+\frac{N}{\pi a^{2}}
\frac{\partial F}{\partial a}=-\frac{\partial U_{a}}{\partial a},
\label{eq23} \\
\zeta _{tt}\equiv \frac{d^{2}\zeta }{dt^{2}} &=&-2\alpha \zeta +2k\beta
f_{1}\sin (k\zeta )\left[ f_{0}e^{-a^{2}k^{2}/4}+f_{1}\cos (k\zeta
)e^{-a^{2}k^{2}}\right]  \nonumber \\
&+&\frac{N}{2\pi a^{2}}\frac{\partial F}{\partial \zeta }=-\frac{\partial
U_{\zeta }}{\partial \zeta },  \label{eq-zeta}
\end{eqnarray}
from where the following effective potentials are identified:
\begin{equation}
\hspace{-1.5cm}U_{a}=\alpha a^{2}+\frac{1}{2a^{2}}+4\beta f_{0}f_{1}\cos
(k\zeta )e^{-a^{2}k^{2}/4}+\beta f_{1}^{2}\cos (2k\zeta )e^{-a^{2}k^{2}}+
\frac{gN}{\sqrt{2\pi }a}-\frac{NF}{\pi a^{2}},  \label{eq-Ua}
\end{equation}
\begin{equation}
U_{\zeta }={\alpha }\zeta ^{2}+2\beta f_{0}f_{1}\cos (k\zeta
)e^{-a^{2}k^{2}/4}+\frac{\beta }{2}f_{1}^{2}\cos (2k\zeta )e^{-a^{2}k^{2}}-
\frac{NF}{2\pi a^{2}}.  \label{eq-Uzeta}
\end{equation}
Frequencies of small oscillations for the width and center of mass can be
obtained from Eqs.~(\ref{eq23}-\ref{eq-Uzeta}). To this end, we set $\zeta
=\zeta _{s}+\delta \zeta ,\delta \zeta \ll \zeta _{s}$, where $\zeta _{s}$
is the fixed point, and $a=a_{s}+\delta a,\delta a\ll a$. By linearizing
Eq.~(\ref{eq23}) in $\delta _{a}$, and Eq.~(\ref{eq-zeta}) in $\delta
_{\zeta }$, respectively, we find
\begin{eqnarray}
\delta a_{tt} &=&\frac{\partial a_{tt}}{\partial a}\delta a+\frac{\partial
a_{tt}}{\partial \zeta }\delta \zeta ,  \label{wa} \\
\delta \zeta _{tt} &=&\frac{\partial \zeta _{tt}}{\partial \zeta }\delta
\zeta +\frac{\partial \zeta _{tt}}{\partial a}\delta a,  \label{wz}
\end{eqnarray}
from where we can derive the corresponding frequencies:
\begin{eqnarray}
\omega _{a}^{2} &=&-\frac{\partial a_{tt}}{\partial a}=2\alpha +\frac{3}{
a_{s}^{4}}+\beta k^{2}f_{0}f_{1}(a_{s}^{2}k^{2}-2)\cos (k\zeta
_{s})e^{-a_{s}^{2}k^{2}/4}  \nonumber \\
&+&2\beta k^{2}f_{1}^{2}(2a_{s}^{2}k^{2}-1)\cos (2k\zeta
_{s})e^{-a_{s}^{2}k^{2}}+\frac{2gN}{\sqrt{2\pi }a_{s}^{3}}  \label{wa2} \\
&-&\frac{6NF}{\pi a^{4}}+\left. \frac{4N}{\pi a^{3}}\frac{\partial F}{
\partial a}\right\vert _{a=a_{s}}-\left. \frac{N}{\pi a^{2}}\frac{\partial
^{2}F}{\partial a^{2}}\right\vert _{a=a_{s}},  \nonumber \\
\omega _{\zeta }^{2} &=&-\frac{\partial \zeta _{tt}}{\partial \zeta }
=2\alpha -2\beta f_{1}k^{2}\left[ f_{0}e^{-a_{s}^{2}k^{2}/4}\cos (k\zeta
_{s})+f_{1}e^{-a_{s}^{2}k^{2}}\cos (2k\zeta _{s})\right]  \label{zeta} \\
&-&\left. \frac{N}{2\pi a^{2}}\frac{\partial ^{2}F}{\partial \zeta ^{2}}
\right\vert _{s}.  \nonumber
\end{eqnarray}

\begin{figure}[tbph]
\centerline{
\includegraphics[width=8.5cm,clip]{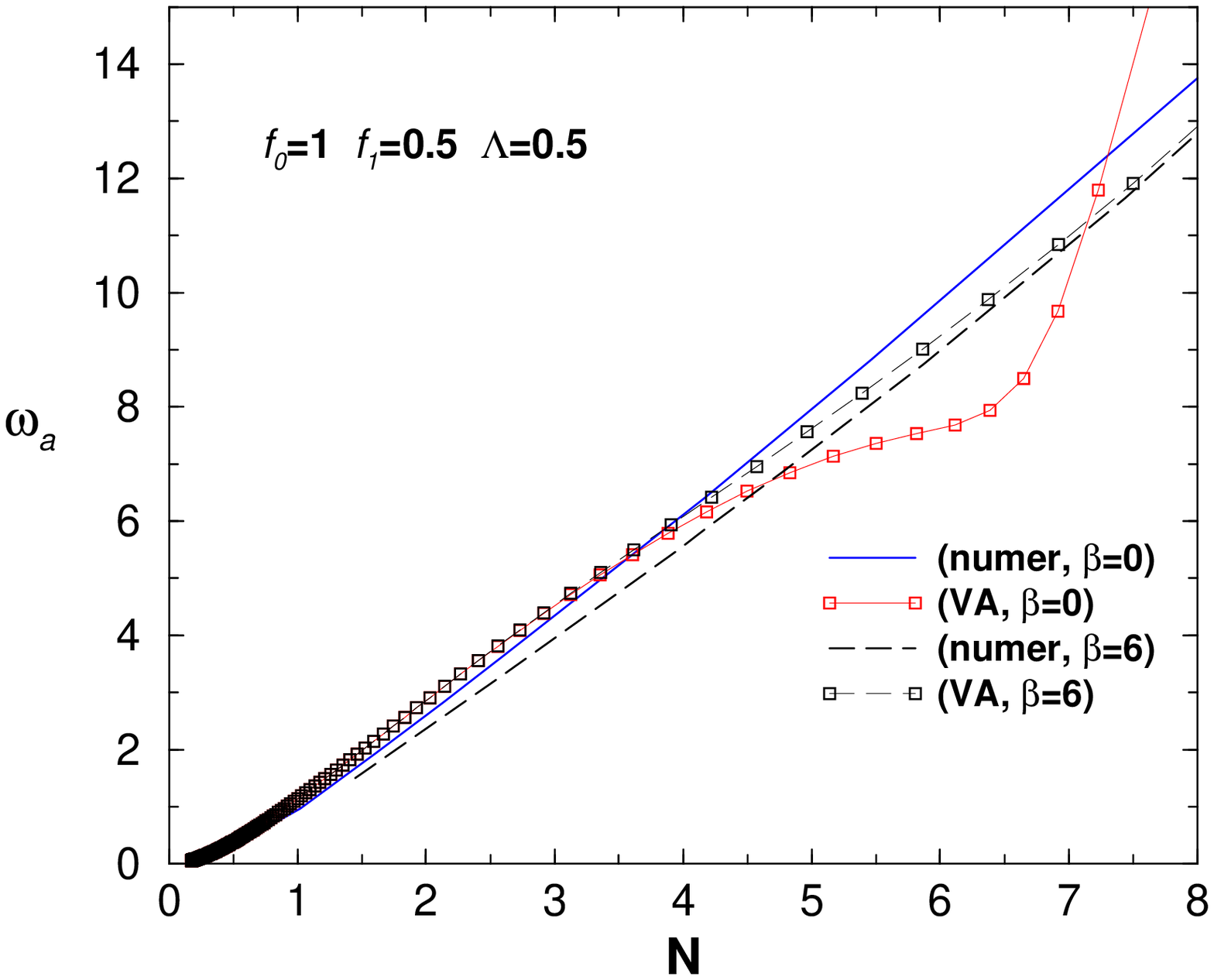}
\hspace{-0.2cm}
\includegraphics[width=8.5cm,clip]{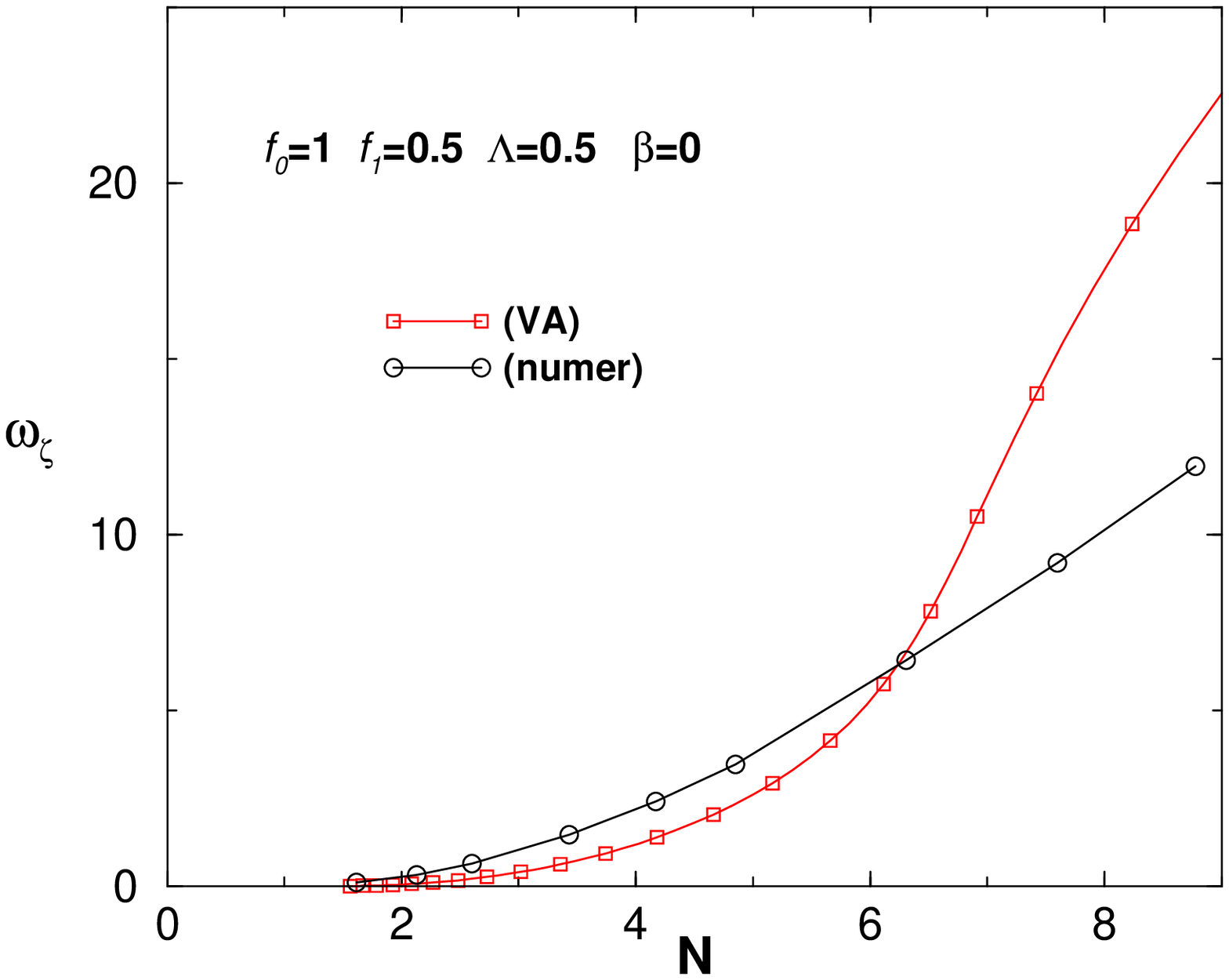}
}
\caption{The left panel: variational results (\protect\ref{wa2}) (red-solid
lines with squares) and numerical solutions (black-solid lines with circles)
for frequencies of small oscillations of perturbed solitons, $\protect\omega 
_{a}$, as functions of $N$. The parameters, as given in the panel, are $
f_{0}=1$, $f_{1}=0.5$, $\Lambda =0.5$, and $\protect\beta =0$. The right
panel: for the same parameters, we show the corresponding variational
results for the frequencies of oscillations of the soliton's center of mass,
$\protect\omega _{\protect\zeta }$ [ as given by Eq. (\protect\ref{zeta})],
compared to the numerical calculations.}
\label{fig09}
\end{figure}

Finally, in Fig.~\ref{fig09} we show results for both frequencies of small
oscillations of the solitons, $\omega _{a}$ and $\omega _{\zeta }$,
comparing the VA predictions to numerical findings. In the left panel we
consider perturbations of their widths, where we observe a good agreement of
the numerical results with the VA. The parameters in this case are $\beta =0$
, $\Lambda =0.5$, $f_{0}=1$, and $f_{1}=0.5$. In the right panel of this
figure, we follow the same procedure by comparing the numerical results with
VA for frequencies related to small oscillations of the center-of-mass, near
$\zeta =0$. Note that the numerical results displayed in the left panel of
Fig.~\ref{fig09} show that the frequency of the width oscillations grows
almost linearly with $N$, in both cases of $\beta =0$ and $\beta =6$. The VA
gives a good description of the results for smaller values of $N$, and the
approximation is improved for larger values of $\beta $. In the case of the
center-of-mass oscillations, displayed only for $\beta =0$, we also observe
that the frequency increases with $N$, although in this case the VA is less
accurate, especially for larger $N$, which is explained by the inadequate
shape of the underlying ansatz (\ref{ans}).

\section{Conclusion}

The objective of this work is to expand the range of settings based on
dipolar BECs, by introducing a model in which atoms or molecules in the free
condensate carry no dipolar moments, but local moments are induced by a
spatially modulated external polarizing field. The DDI (dipole-dipole
interactions) in this setting give rise to an effective nonlocal nonlinear
lattice in the condensate. In the case when the respective interactions are
attractive, they support bright solitons. We have investigated conditions
for the existence of such solitons (including the situation when the
attractive DDI competes with the local repulsion) in the semi-analytical
form, by dint of the VA (variational approximation)\ based on the Gaussian
ansatz. The results were verified by comparison with numerical solutions of
the respective one-dimensional GPE\ (Gross-Pitaevskii equation). The
stability of the soliton families exactly obeys by the VK criterion. The 
dynamics of solitons and interactions
between them, including merger into breathers, were investigated too. In
particular, it was found that the dynamical version of the VA provides for a
good prediction for frequencies of small oscillations of perturbed solitons.

An issue of obvious interest is to extend the analysis reported here for 2D
configurations.

\subsection*{Acknowledgments}

This work was partially supported by the Brazilian agencies Funda\c{c}\~{a}o
de Amparo \`{a} Pesquisa do Estado de S\~{a}o Paulo (FAPESP), Conselho
Nacional de Desenvolvimento Cient\'{\i}fico e Tecnol\'{o}gico (CNPq) and
Coordena\c c\~ao de Aperfei\c coamento de Pessoal de N\'\i vel Superior
(CAPES). BAM was supported, in a part, by the grant from the German-Israeli
Foundation No. I-1024-2.7/2009.

\end{document}